\documentclass[%
 aip,
 amsmath,amssymb,
reprint,%
]{revtex4-1}

\usepackage{graphicx}
\usepackage{dcolumn}
\usepackage{bm}
\usepackage{xcolor}
\usepackage[utf8]{inputenc}
\usepackage[T1]{fontenc}
\usepackage{mathptmx}
\usepackage{eurosym}
\usepackage{ulem}
\usepackage{multirow}
\usepackage{array}
\usepackage{ragged2e}
\newcolumntype{P}[1]{>{\centering\arraybackslash}p{#1}}

\definecolor{giocolor}{RGB}{0, 150, 100}
\newcommand{\santiago}[1]{\textcolor{black}{#1}}

\usepackage{soul}

\begin{document}

\title{Explosive adoption of corrupt behaviors in social systems with higher-order interactions}

\author{Elisa Bretón-Fuertes}
\thanks{These authors contributed equally}
\affiliation{Department of Condensed Matter Physics, University of Zaragoza, 50009 Zaragoza, Spain}

\author{Clara Clemente-Marcuello}
\thanks{These authors contributed equally}
\affiliation{Department of Condensed Matter Physics, University of Zaragoza, 50009 Zaragoza, Spain}

\author{Verónica Sanz-Arqué}
\thanks{These authors contributed equally}
\affiliation{Department of Condensed Matter Physics, University of Zaragoza, 50009 Zaragoza, Spain}

\author{Gabriela Tomás-Delgado}
\thanks{These authors contributed equally}
\affiliation{Department of Condensed Matter Physics, University of Zaragoza, 50009 Zaragoza, Spain}

\author{Santiago Lamata-Otín}
\affiliation{Department of Condensed Matter Physics, University of Zaragoza, 50009 Zaragoza, Spain}
\affiliation{GOTHAM lab, Institute of Biocomputation and Physics of
Complex Systems (BIFI), University of Zaragoza, 50018 Zaragoza, Spain}

\author{Hugo Pérez-Martinez}
\affiliation{Department of Condensed Matter Physics, University of Zaragoza, 50009 Zaragoza, Spain}
\affiliation{GOTHAM lab, Institute of Biocomputation and Physics of
Complex Systems (BIFI), University of Zaragoza, 50018 Zaragoza, Spain}

\author{Jes\'us G\'omez-Garde\~nes}\email{gardenes@unizar.es}
\affiliation{Department of Condensed Matter Physics, University of Zaragoza, 50009 Zaragoza, Spain}
\affiliation{GOTHAM lab, Institute of Biocomputation and Physics of
Complex Systems (BIFI), University of Zaragoza, 50018 Zaragoza, Spain}

\date{\today}

\begin{abstract}
Human behaviors in social systems are often shaped by group pressure and collective norms, especially since the rise of social media platforms. However, in the context of adopting misbehaviors, most existing contagion models rely on pairwise interactions and thus fail to capture group-level dynamics. To fill this gap, we introduce a higher-order extension of the Honesty–Corruption–Ostracism (HCO) model to study the emergence of systemic corruption in populations where individuals interact through group structures. The model incorporates contagion-like transitions mediated by hyperedges of arbitrary order, capturing the influence of peer pressure in group settings. Analytical and numerical results show that higher-order interactions induce discontinuous (explosive) transitions between fully honest and fully corrupt regimes, separated by a bistable phase. This abrupt behavior disappears in the pairwise limit, highlighting the destabilizing effect of group interactions. Furthermore, we establish a general correspondence between our model and broader classes of social contagion dynamics with symmetry breaking, recovering previous results as limiting cases. These findings underscore the critical role of higher-order structure in shaping behavioral adoption processes and the stability of social systems.
\end{abstract}

\maketitle

\textbf{Digital platforms amplify the diffusion of behaviors—both desirable and undesirable, by reinforcing group pressure and shared norms. Within this broad landscape, here we focus on the spread of corrupt conducts. To address the collective nature of these processes, we extend the Honesty–Corruption–Ostracism model to a higher-order framework where interactions take place in groups (hyperedges) of arbitrary size. This formulation captures peer influence at the group level and reveals that such higher-order contagion can trigger abrupt, discontinuous (explosive) shifts between predominantly honest and predominantly corrupt societies, separated by a bistable region. These abrupt transitions disappear in the pairwise limit, underscoring the destabilizing effect of group-mediated interactions. We further map our framework onto more general symmetry-breaking contagion models, recovering known results as limiting cases. Overall, our findings stress that understanding—and mitigating—behavioral shifts on socio-economic platforms requires explicitly accounting for group-level mechanisms, not just individual ties.}

\section{Introduction}
\label{sec:I}

Higher-order systems, which account for group-level rather than pairwise interactions, provide a refined framework to represent complex social systems \cite{battiston2020networks,bick2023higher,majhi2022dynamics}. Such group interactions, formally described using hypergraphs, have revealed a variety of novel collective phenomena \cite{battiston2021physics}, especially in the realm of social dynamics, in which abrupt (explosive) phase transitions have turned ubiquitous when agents interact in groups \cite{iacopini2019simplicial,landry2020effect,st2022influential,Civilini2024explosive,ferraz2024contagion,malizia2025hyperedge}. 
\santiago{Furthermore, higher-order frameworks have also been employed to study the evolutionary dynamics of strategies and the spread of information, both in static and adaptive settings \cite{alvarez2021evolutionary,dong2025adaptive}.}
Importantly, contemporary socio-economic platforms (e.g., large-scale social media) intensify these higher-order effects by structuring interactions in groups, communities, and channels rather than in isolated dyads, further motivating models that move beyond pairwise assumptions \cite{vosoughi2018spread,bakshy2012role,aral2009distinguishing}.
\medskip

In its turn, social collective phenomena are ultimately rooted in behavioral changes, which not only shape individual trajectories but also drive emergent patterns at the population level \cite{castellano2009statistical}. Behavior adoption processes underpin a wide range of social dynamics, such as the diffusion of cultural norms \cite{axelrod1997dissemination}, technological innovation \cite{granovetter1978threshold,centola2010spread}, rumor spreading \cite{daley1964epidemics,maki1973mathematical}, \santiago{opinion polarization \cite{starnini2025opinion,perez2025social},} or the emergence of cooperation \cite{axelrod1981evolution}, and are known to respond nonlinearly to group influences. On digital platforms these processes are further channeled and amplified by group-based features (e.g., group chats, forums, and curated communities), which act as natural drivers for multi-individual influence.
\medskip

Among behaviors with strong societal impact, systemic corruption remains a major global concern, with estimated costs exceeding $5\%$ of the World Gross Domestic Product \cite{un_corruption_costs_2018}, and consistently listed among the top public issues worldwide \cite{ipsos2024}. 
\santiago{Besides, recent analyses have revealed consistent structural patterns and demographic disparities in real corruption systems, across different contexts and scales \cite{martins2022universality,pessa2025structural}.}
Corrupt practices arise in diverse contexts—social, political, economic—and often involve group pressures or complicity mechanisms. 
In platform-mediated environments, such pressures may be reinforced by the visibility of group norms and coordinated actions, reinforcing the need for frameworks that explicitly incorporate higher-order interactions.

\medskip

Mathematical modeling efforts have addressed corruption from different perspectives. A large body of work relies on game-theoretic approaches \cite{lee2019social,kolokoltsov2012nonlinear,kolokoltsov2017mean,lee2015games,lee2017games,verma2015bribe,verma2017bribery,verma2018bribery,von2007theory}, where behaviors are treated as strategic choices aimed at maximizing payoffs. 
More recently, compartmental models have been introduced to describe corruption as a contagion-like process \cite{lu2020norm,bauza2020fear,perez2022emergence,lu2023dynamics}, inspired by epidemic dynamics \cite{kermack1927contribution,barrat2008dynamical,boccara2010modeling}. In particular, the Honesty–Corruption–Ostracism (HCO) model \cite{lu2020norm} considers individuals as honest (H), corrupt (C), or ostracized (O), with transitions mediated by interactions: corrupt agents can corrupt honest ones, while honest agents can denounce corrupt individuals. However, the original HCO framework assumes only pairwise interactions, thus neglecting the influence of group dynamics in shaping behavior.
\medskip

In this article, we extend the HCO model to incorporate higher-order contagion mechanisms, yielding the Higher-Order HCO (HO-HCO) model. This extension allows us to investigate how group-level interactions affect the onset of systemic corruption and whether they give rise to discontinuous (explosive) transitions. Our objectives are twofold: (\textit{i}) to characterize the impact of higher-order interactions on corruption dynamics, and (\textit{ii}) to examine how these findings relate to recent insights on collective behavior adoption in social systems \cite{iacopini2019simplicial}.
\medskip

The structure of the paper is as follows. In Section~\ref{sec:II}, we introduce the HO-HCO model and its mean-field formulation. In Section~\ref{sec:III}, we show that higher-order interactions induce explosive transitions between honest and corrupt states. In Section~\ref{sec:IV}, we connect this behavior with general models of behavioral adoption. Finally, we summarize and discuss the implications of our findings in Section~\ref{sec:V}.

\begin{figure*}[t!]
\centering\includegraphics[width=0.88\linewidth]{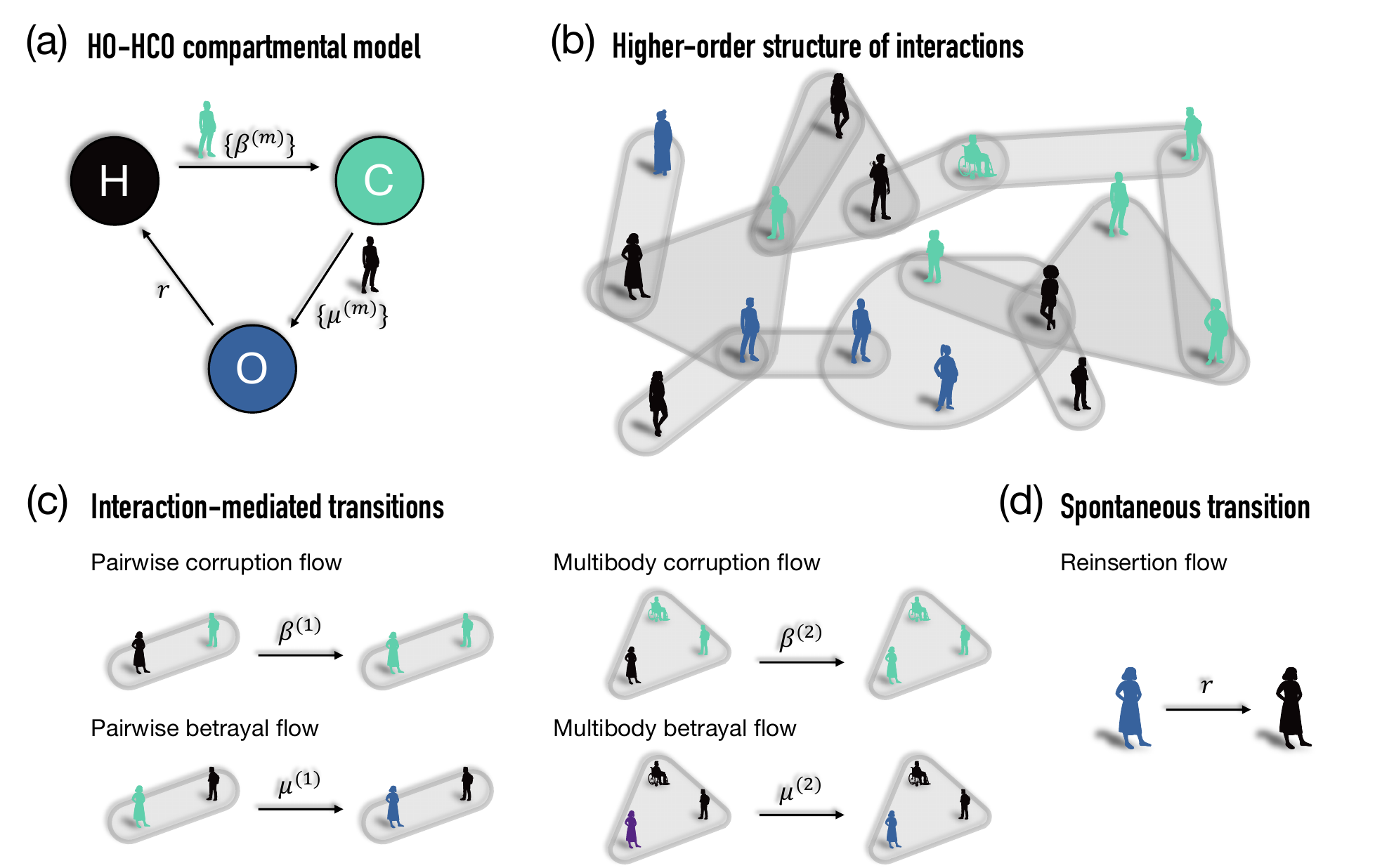}
\caption{\textbf{Schematic illustration of the HO-HCO dynamical model}. \santiago{The framework combines pairwise and group-mediated corruption and betrayal processes.} Panel (a) depicts the compartmental structure, where honest individuals can transition to the corrupt compartment according to the probabilities $\{\beta^{(m)}\}$, corrupt individuals can transition towards the ostracism compartments according to the probabilities $\{\mu^{(m)}\}$, and those in ostracism can be reinserted into society with probability $\mu$. Note that $m$ is the order of interaction. Panel (b) represents the higher-order structure of interactions. Panel (c) illustrates the interaction-mediated transition, i.e. the corruption and betrayal flows. Panel (d) illustrates the spontaneous reinsertion flow.}
\label{fig:P1}
\end{figure*}

\bigskip

\section{Group dynamics of corruption}
\label{sec:II}

As anticipated above, the adoption of corrupt behavior can be framed as a contagion process: exposure to corrupt peers increases the likelihood that an honest individual becomes corrupt. Symmetrically, betrayal also spreads through social influence, as honest individuals who interact with wrongdoers may report their malpractice \cite{lu2023dynamics}. Depending on the nature of social interactions, corruption and betrayal processes occur through pairwise interactions or through groups of more than two individuals. For the sake of generality in the model definition, in this section we define the order of an interaction, $m$ as the number of individuals involved in the interaction minus one. Moreover, we set $M$ to be the maximum order of interactions.
\medskip

\subsection{Higher-order corruption model}

Formally, we build a compartmental model with three states or compartments: Honest (H), Corrupt (C) and Ostracism (O) (see Fig. \ref{fig:P1}.a). The transitions between these states are depicted in Fig. \ref{fig:P1}.b-d and are explained below. Individuals designated as Honest (H) are those adopters of the well-established norms and laws of society, and can potentially become corrupt. The adoption of corrupt behavior can occur through direct contact with Corrupt (C) agents. Following the rules for social contagion established by Iacopini et al. \cite{iacopini2019simplicial}, an honest individual may become corrupt due to its interaction with a group of $m+1$ individuals if the $m$ other individuals involved in the interaction are corrupt, with a probability $\beta^{(m)}$. Upon adoption, Honest (H) agents transit into the Corrupt (C) state, so that they can induce honest individuals to violate the norms. In its turn, Corrupt (C) agents can be betrayed by interacting with honest individuals. Analogously to the adoption process, a corrupt individual is betrayed due to its interaction with a group of $m+1$ individuals if the other $m$ individuals involved in the interaction are honest, with a probability $\mu^{(m)}$. Those betrayed individuals join the Ostracism (O) state, which can be understood as being out of society in punishment for their actions. Those individuals in the Ostracism (O) reinsert into the honest population after an average of $r^{-1}$ time units, without requiring interaction with other agents.
According to the former description, the HO-HCO model comprises three states and $2M+1$ parameters. 
\medskip

As illustrated in Fig. \ref{fig:P1}.c, there are two transitions associated to interactions between individuals. Namely: $H\rightarrow C$ and $C\rightarrow O$. These transitions depend on the structure of the society, which here is embodied by an hypergraph (see Fig. \ref{fig:P1}.b). Mathematically, an hypergraph $\mathcal{H}=\mathcal{H}(\mathcal{N},\mathcal{E})$ is a pair of two sets: $\mathcal{N}$, composed of $N=|\mathcal{N}|$ nodes (here representing the individuals), and $\mathcal{E}$, that contains a number $E=|\mathcal{E}|$ of hyperedges (here representing the groups). An hyperedge $e\in\mathcal{E}$ of order $m$, i.e. an $m$-hyperedge, is defined as a subset of $m+1$ nodes in $\mathcal{N}$, being $m=|e|-1$. Therefore, pairwise interactions between two nodes correspond to hyperedges of order 1, interactions among three nodes are represented by hyperedges of order 2, and larger groups are similarly represented by higher-order hyperedges. Each node $i$ belongs to a set of hyperedges of order $m$, $\mathcal{E}_i^{(m)}$. Therefore, we can define the generalized degree of node $i$, namely $k_i^{(m)}$, as the cardinality of that set, i.e. $k_i^{(m)}=|\mathcal{E}_i^{(m)}|$.

\subsection{Mean field dynamical equations}

\begin{figure*}[t!]
\centering\includegraphics[width=0.83\linewidth]{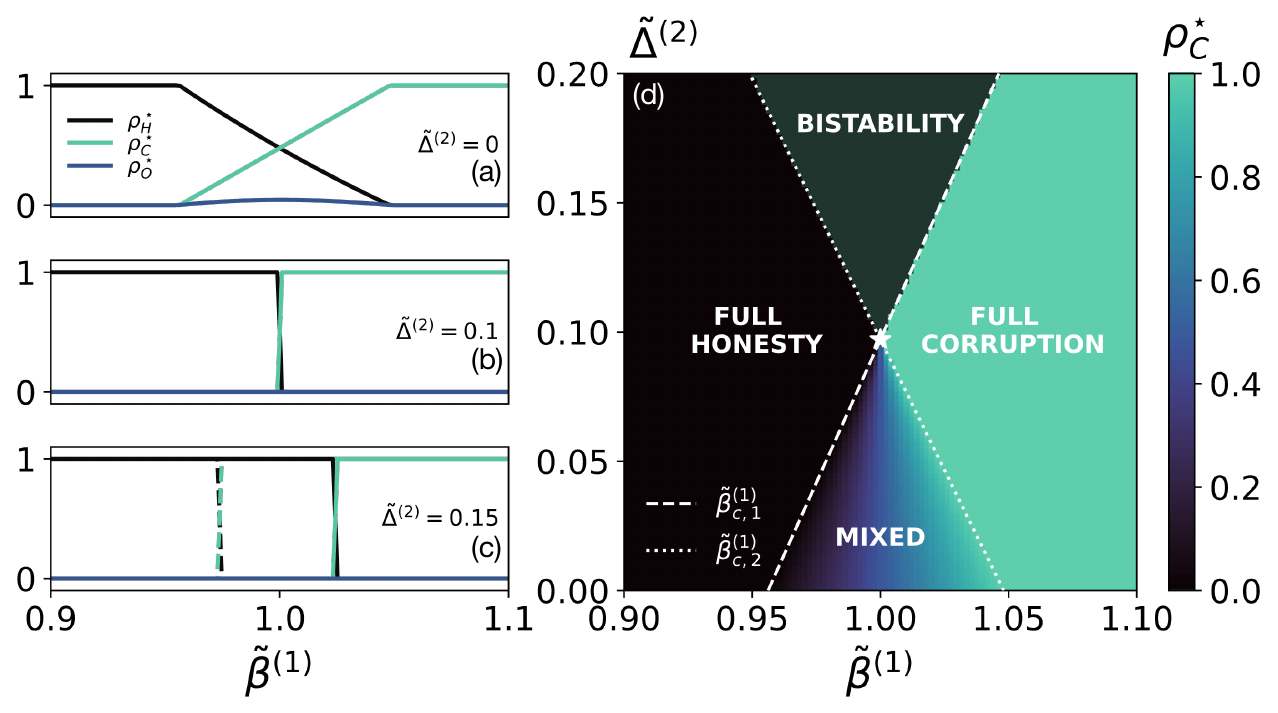}
\caption{\textbf{Effect of higher-order interactions on corruption dynamics}. \santiago{Higher-order interactions suppress the mixed phase and trigger explosive shifts between honest and corrupt societies.} (a) Phase diagram for the HO-HCO model. Four phases emerge as a function of $\beta^{(1)}/\mu^{(1)}$ and of the higher-order relevance $\Delta^{(2)}/\mu^{(1)}$: a full-honesty phase with $\rho_H^{\star}=1$, a full-corruption phase with $\rho_C^{\star}=1$, a mixed phase with $\rho_H^{\star}\neq0$, $\rho_C^{\star}\neq0$ and $\rho_O^{\star}\neq0$, and a bistability phase, where the stationary state (either $\rho_H^{\star}=1$ or $\rho_C^{\star}=1$) depends on the initial conditions. Note that the theoretical expressions given by Eqs. (\ref{eq:8})-(\ref{eq:9}) delimit the phases. (b) Three cuts of the diagram: For $\Delta^{(2)}=0$ the pairwise phenomenology is recovered \cite{lu2020norm}. For $\Delta^{(2)}=\Delta^{(2)}_{\star}$ there is an abrupt transition without bistability. For $\Delta^{(2)}=0.0015$ there is an explosive transition with bistability region (continuous and dashed lines correspond to $\rho_C(0)=0.01$ and $\rho_C(0)=0.99$ initial conditions).
In all panels $k^{(1)}=k^{(2)}=10$, $\mu^{(1)}=0.01$, $\beta^{(2)}=\mu^{(2)}=\Delta/2$ and $r=0.5$. Note that $\tilde\beta^{(1)}=\beta^{(1)}/\mu^{(1)}$ and $\tilde\Delta^{(2)}=\Delta^{(2)}/\mu^{(1)}$.}
\label{fig:P2}
\end{figure*}

The HO-HCO dynamics, can be studied under a microscopic Markovian time-discrete formulation \cite{granell2024probabilistic}. For simplicity, we assume that all individuals are equivalent and well mixed. This leads to a mean field description in which the relevant variables are the probabilities of being honest, $\rho_H(t)$, corrupt, $\rho_C(t)$, and in the ostracism state, $\rho_O(t)$, together with the average generalized degrees $k^{(m)}$ ($m=1,...,M$). The evolution of these probabilities is given by:
\begin{eqnarray}
    \rho_C(t+1)&=&\left[1-\Pi^{C\rightarrow O}(t)\right]\rho_C(t)+\Pi^{H\rightarrow C}(t)\rho_H(t),
    \label{eq:1}\\
    \rho_H(t+1)&=&\left[1-\Pi^{H\rightarrow C}(t)\right]\rho_H(t)+r\rho_O(t),
    \label{eq:2}
\end{eqnarray}
and since the sum of probabilities at some time $t$ must be equal to 1 we have:
\begin{eqnarray}
    \rho_O(t)=1-\rho_C(t)-\rho_H(t).
    \label{eq:3}
\end{eqnarray}
Now let us describe the terms associated to interaction-mediated transitions. In the former set of equations, the corruption and betrayal flows are captured by the probabilities $\Pi^{H\rightarrow C}(t)$ and $\Pi^{C\rightarrow O}(t)$, that read as follows:
\begin{eqnarray}
    \Pi^{H\rightarrow C}(t)&=&1-\prod_{m=1}^M \left[1-\beta^{(m)}\rho_C^m(t)\right]^{k^{(m)}},
    \label{eq:4}\\
    \Pi^{C\rightarrow O}(t)&=&1-\prod_{m=1}^M \left[1-\mu^{(m)}\rho_H^{m}(t)\right]^{k^{(m)}}.
    \label{eq:5}
\end{eqnarray}
Their functional form is inspired by the effective transition probabilities arising from stochastic dynamics \cite{lu2020norm,bauza2020fear,perez2022emergence}. The effective probability, $P$, of an individual transitioning to the corrupt (ostracism) state is complementary to the effective probability of not transitioning, $\bar P$, i.e. $P=1-\bar P$. This latter probability ($\bar P$) is the product of the probabilities of events leading to transition, namely the contact with each of the $k^{(m)}$ groups of every order $m=1,...,M$, times the corresponding transition rate. In Supplementary Fig. 1, we compare the integration of Eqs. (\ref{eq:1})-(\ref{eq:5}) with the outcome of stochastic Monte Carlo simulations, showcasing the goodness of the framework.

\section{Explosive onset of corruption}
\label{sec:III}

The pairwise HCO model can lead to three dynamical outcomes \cite{lu2020norm}: full-honesty, where all individuals are honest; full-corruption, where all individuals are corrupt; and a mixed state, where there is a dynamical equilibrium with individuals in each of the three states (honest, corrupt, and ostracism).
In this section, we explore the nature of the transitions between the three aforementioned states, both in the absence and presence of higher-order interactions. To better understand the influence of groups, we introduce $\Delta^{(2)}=\beta^{(2)}+\mu^{(2)}$ as control parameter of higher-order relevance. Moreover, we set  $\beta^{(2)}=\mu^{(2)}=\Delta^{(2)}/2$, $k^{(1)}=k^{(2)}=10$, $\mu^{(1)}=0.01$ and $r=0.5$ throughout this section. This leaves the corruption strength ($\beta^{(1)}$) and the higher-order relevance ($\Delta^{(2)}$) as control parameters. Note that for clarity purposes, we are also utilizing in the figures the rescaled version of the parameters $\tilde\beta^{(1)}=\beta^{(1)}/\mu^{(1)}$ and $\tilde\Delta^{(2)}=\Delta^{(2)}/\mu^{(1)}$.
\medskip

In Fig. \ref{fig:P2}.a-c we display the stationary fractions of corrupt, honest and out of society individuals, i.e. $\rho^{\star}_x=\text{lim}_{t\rightarrow \infty}\rho_X(t)$, with $X=H,C,O$. In the absence of higher-order interactions ($\tilde\Delta^{(2)}=0$), as $\tilde\beta^{(1)}$ increases the system smoothly transitions from the full-honesty state to a mixed state, where there is a fraction of the society in each compartment (see Fig. \ref{fig:P2}.a). For even larger values of $\tilde\beta^{(1)}$, the system undergoes another continuous transition, now from the mixed state to the full-corruption scenario. 
From Fig. \ref{fig:P2}.b we observe that, as the relevance of higher-order interactions increases ($\tilde\Delta^{(2)}=0.1$), the region of parameters leading to a mixed society disappears, yielding an abrupt shift between the full-honesty and full-corruption stationary states. Finally, when considering in Fig. \ref{fig:P2}.c a large relevance of higher-order interactions ($\tilde\Delta^{(2)}=0.15$), we observe an explosive transition between the full-honesty and full-corruption stationary states, with no mixed scenario in between. Moreover, in this case, there is also a bistability region, pinpointing that the precise stationary state reached depends on the set of initial conditions. To better perceive the change in behavior induced by higher-order interactions, we show in Fig. \ref{fig:P2}.d the full phase diagram in the ($\tilde\beta^{(1)}-\tilde\Delta^{(2)}$)-space by representing $\rho_C^{\star}$. There, four regions arise: full-honesty, full-corruption, mixed, and bistability. Importantly, the bistable regime only arises above a certain relevance of higher-order interactions, namely when $\Delta^{(2)}>\Delta^{(2)}_{\star}$.
\medskip

In order to understand the transitions between regions, we perform a stability analysis around the full-honesty and full-corruption stable states (see Supplementary Eqs. (S.1)-(S.14)) by evaluating the Jacobian of the system of Eqs. (\ref{eq:1})-(\ref{eq:5}), after transforming this system into its continuous time version. 
The general stability condition for the full-honesty state ($\rho_H=1$, $\rho_C=0$) reads
\begin{eqnarray}
    k^{(1)}\beta^{(1)}+\prod_{m=1}^M\left[1-\mu^{(m)}\right]^{k^{(m)}}<1\;,
    \label{eq:6}
\end{eqnarray}
while the general condition for the full-corruption state  ($\rho_H=0$, $\rho_C=1$) is
\begin{eqnarray}
    k^{(1)}\mu^{(1)}+\prod_{m=1}^M\left[1-\beta^{(m)}\right]^{k^{(m)}}<1\;.
    \label{eq:7}
\end{eqnarray}
Now, based upon Eqs. (\ref{eq:6})-(\ref{eq:7}) and restricting group interactions to three-body ones, $M=2$, we can analytically derive the boundaries representing the transitions onsets in Fig. \ref{fig:P2}.d:
\begin{eqnarray}
    \beta^{(1)}_{c,1}&=&\frac{1-\left[1-\mu^{(1)}\right]^{k^{(1)}}\left[1-\mu^{(2)}\right]^{k^{(2)}}}{k^{(1)}},
    \label{eq:8}\\
    \beta^{(1)}_{c,2}&=&1-\left\{\frac{1-k^{(1)}\mu^{(1)}}{\left[1-\beta^{(2)}\right]^{k^{(2)}}}\right\}^{\frac{1}{k^{(1)}}}\;,
    \label{eq:9}
\end{eqnarray}
being $\beta^{(1)}_{c,1}$ the boundary of the full-honesty phase and $\beta^{(1)}_{c,1}$ the boundary of the full-corruption phase.
\medskip

Remarkably, we can also derive the minimum relevance of higher-order interactions required in order to obtain an explosive transition between full-honesty and full-corruption states. This quantity is usually known as the tricritical point ($\beta^{(1)}_{\star},\Delta_{\star}^{(2)}$) and by imposing that $\beta^{(1)}_{\star}=\beta^{(1)}_{c,1}=\beta^{(1)}_{c,2}$ we obtain its implicit expression as:
\begin{widetext}    
\begin{eqnarray}
    \Delta_{\star}=2\left\{1-\left\{
    \frac{
    \left[
    1+k^{(1)}\left(
    \left[
    \frac{1-k^{(1)}\mu^{(1)}}{\left(1-\frac{\Delta_{\star}}{2}\right)^{k^{(2)}}}
    \right]^{\frac{1}{k^{(1)}}}
    -1\right)\right]^{\frac{1}{k^{(2)}}}}
    {\left[1-\mu^{(1)}\right]^{\frac{k^{(1)}}{k^{(2)}}}}
    \right\}\right\}.
    \label{eq:10}
\end{eqnarray}
\end{widetext}
The outcome of the analytical expressions in Eqs. (\ref{eq:8})-(\ref{eq:10}) is depicted in Fig. \ref{fig:P2}, and agrees well with the numerical iteration of Eqs. (\ref{eq:1})-(\ref{eq:5}).
\medskip

It is also worth noting that the inclusion of higher-order interactions forbids the mixed scenario if $\Delta^{(2)}>\Delta^{(2)}_{\star}$. From a dynamical point of view, the ostracism compartment only induces a delay in the reinsertion process. Moreover, the spontaneous nature of the flow $O\rightarrow H$ means that, if there are no corrupt individuals, it is unavoidable that those individuals in the $O$ state will be reinserted into society. Thus, the additional interaction‑mediated pathways accelerate the dynamics and make it easier for either the honest or the corrupt compartment to be emptied. Consequently, the system bypasses the mixed stationary state.

\section{Explosive shift between competing behaviors in social systems}
\label{sec:IV}

\subsection{The HO-HC model}

\begin{figure}[t!]
\centering\includegraphics[width=0.95\linewidth]{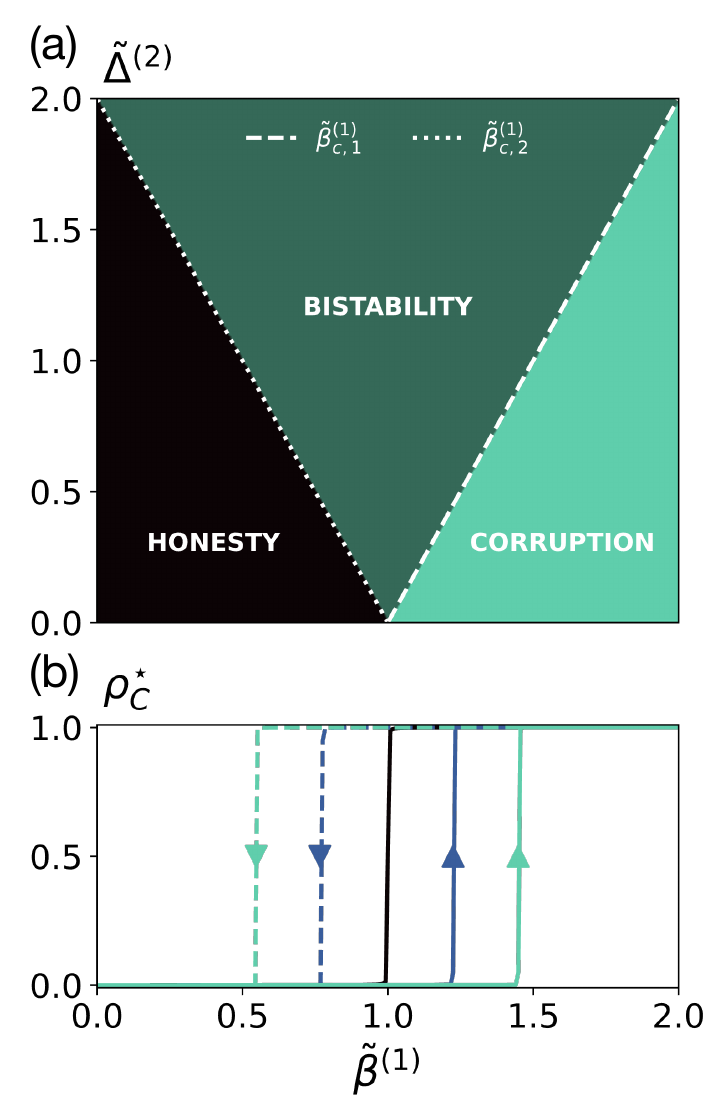}
\caption{\textbf{Explosive shift between honest and corrupt societies}. \santiago{Higher-order effects broaden the bistability region, where both honest and corrupt societies are stable.} (a) Phase diagram for the HO-HC model. Three phases emerge as a function of $\beta^{(1)}/\mu^{(1)}$ and of the higher-order relevance $\Delta^{(2)}/\mu^{(1)}$: a full-honesty phase with $\rho^{\star}=0$, a full-corruption phase with $\rho^{\star}=1$, and a bistability phase, where the stationary state (either $\rho^{\star}=0$ or $\rho^{\star}=1$) depends on the initial conditions. Note that the theoretical expressions given by Eqs. (\ref{eq:8})-(\ref{eq:9}) delimit the phases. (b) Three cuts of the diagram: For $\Delta^{(2)}=0$ is an abrupt transition without bistability, and for $\Delta^{(2)}>0$ there is an explosive transition with bistability region (continuous and dashed lines correspond to $\rho(0)=0.01$ and $\rho(0)=0.99$ initial conditions).
In both panels $k^{(1)}=k^{(2)}=10$, $\beta^{(2)}=\mu^{(2)}=\Delta/2$ and $\mu^{(1)}=0.01$. Note that $\tilde\beta^{(1)}=\beta^{(1)}/\mu^{(1)}$ and $\tilde\Delta^{(2)}=\Delta^{(2)}/\mu^{(1)}$.}
\label{fig:P3}
\end{figure}

\begin{figure*}[t!]
\centering\includegraphics[width=0.9\linewidth]{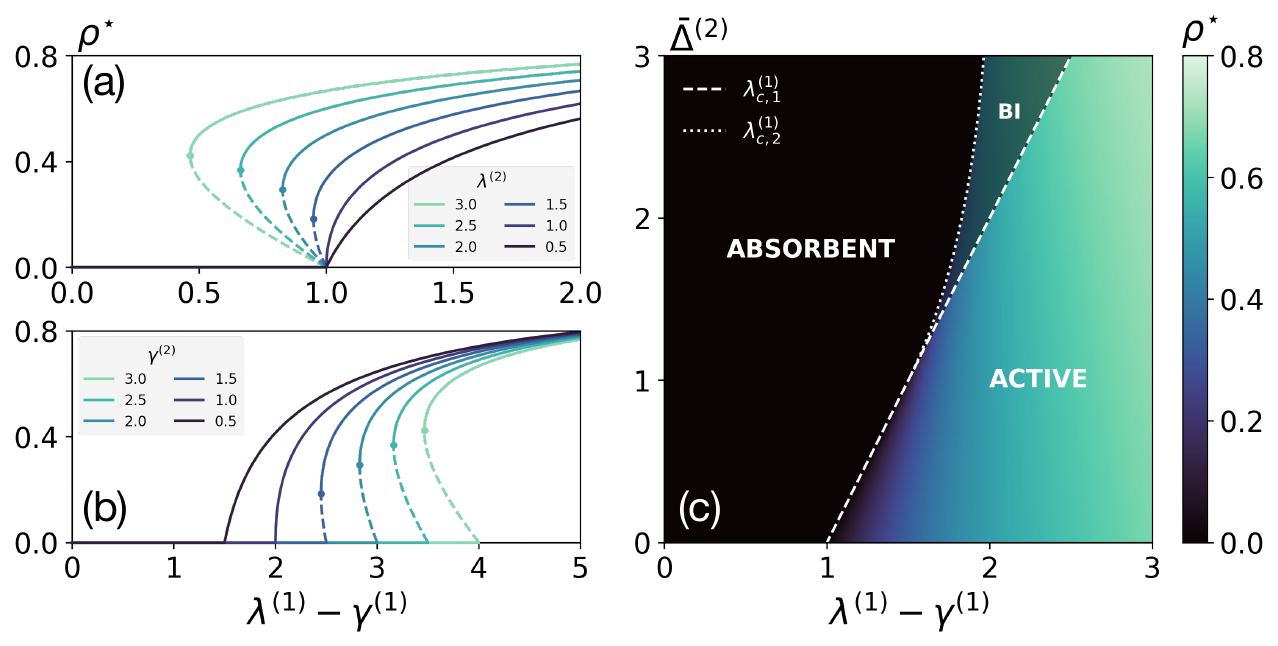}
\caption{\textbf{Explosive adoption of behaviors in social systems}.
\santiago{In a general behavior adoption framework, higher-order interactions transform the transition to the active state from continuous to discontinuous.}
(a)-(b) Stationary solutions $\rho^{\star}$ given by Eq. (\ref{eq:19}) are represented in terms of the ratio of rescaled pairwise parameters ($\lambda^{(1)}-\gamma^{(1)}$). Each curve corresponds to a different value of $\lambda^{(2)}$ or $\gamma^{(2)}$ respectively.
In both panels (a)-(b) the continuous and dashed lines depict the stable and unstable branches respectively, and increasing the value of $\lambda^{(2)}$ and $\gamma^{(2)}$ changes the nature of the transition, which becomes discontinuous. In panel (a) we fix $\gamma^{(2)}=0$ and $\gamma^{(1)}=1$, and in panel (b) we fix $\lambda^{(2)}=0$ and $\gamma^{(1)}=1$.
(c) Phase diagram for the general framework. Three phases emerge as a function of $\lambda^{(1)}-\gamma^{(1)}$ and of the higher-order relevance $\Delta^{(2)}$: an absorbent phase with $\rho^{\star}=0$, an active phase with $\rho^{\star}\neq0$, and a bistability phase, where the stationary state (either $\rho^{\star}=0$ or $\rho^{\star}\neq0$) depends on the initial conditions. Note that the theoretical expressions given by Eqs. (\ref{eq:20})-(\ref{eq:21}) delimit the phases. 
In panel (c), $\gamma^{(1)}=1$ and $\beta^{(2)}=\mu^{(2)}=\Delta/2$. Note that $\bar\Delta^{(2)}=k^{(2)}\Delta^{(2)}/\mu^{(0)}$ }
\label{fig:P4}
\end{figure*}

As explained above, in the presence of strong higher-order interactions ($\Delta^{(2)}>\Delta^{(2)}_{\star}$), the stationary relevance of Ostracism is lost. Therefore, the interaction between honest and corrupt populations reduces to a peer‑pressure contest in which each side aims to convert the other. To better understand the explosive shifts, we assume that corrupt individuals are immediately reinserted into society without punishment ($r\rightarrow\infty$). Under this assumption (see Supplementary Eqs. (1)-(5)) the system dynamics can be described by a single differential equation that parametrizes the fraction of corrupt individuals, hereafter $\rho(t)\equiv\rho_C(t)$:
\begin{eqnarray}
     \dot\rho(t)&=&\beta^{(1)}k^{(1)}\rho(t)(1-\rho(t))+\beta^{(2)}k^{(2)}\rho^2(t)(1-\rho(t))\\
    &&-\mu^{(1)}k^{(1)}(1-\rho(t))\rho(t)-\mu^{(2)}k^{(2)}(1-\rho(t))^2\rho(t).\nonumber
    \label{eq:11}
\end{eqnarray}
Note that, provided the sum of probabilities must remain equal to 1, the fraction of honest individuals is now $\rho_H(t)=1-\rho(t)$. In Eq. (\ref{eq:11}) we have also simplified the transition probabilities by assuming that interaction events are independent and therefore the intersection of probabilities inspiring Eqs. (\ref{eq:4})-(\ref{eq:5}) is the null space. 
\medskip

Under the former assumptions and by imposing $\dot\rho=0$, we are able to derive that the dynamical system has three stationary solutions:
\begin{eqnarray}
    \rho^{\star}_1&=&0,
    \label{eq:12}\\
    \rho^{\star}_2 &=& \frac{\mu^{(1)} k^{(1)} - \beta^{(1)} k^{(1)} + \mu^{(2)} k^{(2)}}{\beta^{(2)} k^{(2)} + \mu^{(2)} k^{(2)}},
    \label{eq:13}\\ 
    \rho^{\star}_3&=&1,
    \label{eq:14}
\end{eqnarray}
which correspond to the fully-honest stable state ($\rho_1^{\star}$), the unstable state that acts as the border between the basins of attraction ($\rho_2^{\star}$), and the fully-corrupt stable state ($\rho_3^{\star}$) respectively. In Fig. \ref{fig:P3}.a we represent the phase diagram resulting from Eq. (\ref{eq:11}) in the ($\tilde\beta^{(1)},\tilde\Delta^{(2)}$)-space. As anticipated, three regions emerge: fully-honest, fully-corrupt, and bistable.
In this case, the critical values of $\beta$ delimiting the basins of attraction of the fully-honest and fully-corrupt states read $\beta_{c,1}^{(1)}=\left(\mu^{(1)} k^{(1)} + \mu^{(2)} k^{(2)}\right)/k^{(1)}$ and $\beta_{c,2}^{(1)}=\left(\mu^{(1)} k^{(1)} - \beta^{(2)} k^{(2)}\right)/k^{(1)}$.
\medskip

Interestingly, for every relevance of higher-order interactions, the system always displays bistability for certain values of $\beta^{(1)}$, since $\beta_{c,1}^{(1)}>\beta_{c,2}^{(1)}$. Moreover, the width of this region, $\Lambda(\beta^{(1)})=\beta_{c,1}^{(1)}-\beta_{c,2}^{(1)}$, is linearly proportional to the relevance of higher-order interactions:
\begin{eqnarray}
    \Lambda(\beta^{(1)})=\frac{k^{(2)}}{k^{(1)}}\left(\mu^{(2)}+\beta^{(2)}\right)=\frac{k^{(2)}}{k^{(1)}}\Delta^{(2)}.
    \label{eq:15}
\end{eqnarray}
The cuts of the diagram in Fig. \ref{fig:P3}.b exemplify how the bistability region is more prominent as the relevance of higher-order interactions is augmented.
Remarkably, while this implies that the volatility of the dynamics is higher, it also pinpoints that, when starting from a mostly honest society (below the critical value of Eq. (\ref{eq:13})), the range of parameters leading to the fully-honest state is wider.

\subsection{General higher-order framework for the adoption of behaviors in social systems}

In the previous subsection, we have characterized the competition of two dynamical processes where individuals require interaction with adopters for transitioning. However, depending on the nature of the adopted behavior, the transitions between states can be either interaction-mediated and/or spontaneous. Therefore, Eq. (\ref{eq:11}) can be generalized to incorporate every order of interaction and to include also spontaneous transition rates, $\beta^{(0)}$ and $\mu^{(0)}$:
\begin{eqnarray}
    \dot\rho(t)&=&\sum_{m=0}^M\beta^{(m)}k^{(m)}\rho^m(t)\left(1-\rho(t)\right)\nonumber \\
    &&-\sum_{m=0}^M\mu^{(m)}k^{(m)}\left(1-\rho(t)\right)^m\rho(t),
    \label{eq:16}
\end{eqnarray}
where we acknowledge that $m=0$ involves only the node itself and therefore $k^{(0)}=1$. 
\medskip

When two behaviors compete for prevalence within a society, it can be argued that there is one basal behavior for the individuals, i.e. that they tend to adopt one of the behaviors in isolation (symmetry braking). Particularly, in the honest-corrupt interplay, it could be argued that individuals naturally tend to follow the rules. Therefore, we can assume that $\beta^{(0)}=0$ and $\mu^{(0)}\neq0$ and, for $M=2$ and $\beta^{(0)}=0$, Eq. (\ref{eq:16}) reads as follows:
\begin{eqnarray}
    \dot\rho(t)&=&\beta^{(1)}k^{(1)}\rho(t)(1-\rho(t))+\beta^{(2)}k^{(2)}\rho^2(t)(1-\rho(t))\nonumber \\
    &&-\mu^{(1)}k^{(1)}(1-\rho(t))\rho(t)-\mu^{(2)}k^{(2)}(1-\rho(t))^2\rho(t)\nonumber\\
    &&-\mu^{(0)}\rho(t),
    \label{eq:18}
\end{eqnarray}
which is a cubic equation in $\rho(t)$ that can be analytically solved in the stationary state ($\dot\rho=0$). By solving it we note that the system keeps the usual absorbent state, $\rho_1^{\star}=0$, as a solution and the other two equilibria of the equations are the roots of a second-order polynomial equation. Absorbing the generalized degrees and rescaling every ratio by $\mu^{(0)}$, i.e. $\lambda^{(m)}=\beta^{(m)}k^{(m)}/\mu^{(0)}$ and $\gamma^{(m)}=\mu^{(m)}k^{(m)}/\mu^{(0)}$, we can express the non-trivial solutions of Eq. (\ref{eq:18}) as:
\begin{widetext}
\begin{eqnarray}
    \rho^{\star}_{2\,\pm}=\frac{\gamma^{(1)}-\lambda^{(1)}+\lambda^{(2)}+2\gamma^{(2)}\pm\sqrt{\left(\gamma^{(1)}-\lambda^{(1)}+\lambda^{(2)}+2\gamma^{(2)}\right)^2-4\left(\lambda^{(2)}+\gamma^{(2)}\right)\left(1-\lambda^{(1)}+\gamma^{(1)}+\gamma^{(2)}\right)}}{2\left(\lambda^{(2)}+\gamma^{(2)}\right)}.
    \label{eq:19}
\end{eqnarray}
\end{widetext}
\medskip

In light of the three solutions, namely, $\rho_1^{\star}$, $\rho_{2\,\pm}^{\star}$, three distinct scenarios emerge. If none of the solutions of Eq. (\ref{eq:19}) is defined as positive, the absorbent state $\rho_1^{\star}=0$ is the stable solution. In that case, there is a full-honest society. Conversely, if there is only one defined positive solution of Eq. (\ref{eq:19}), the absorbent state $\rho_1^{\star}=0$ becomes unstable, and there is a steady stationary solution, where a fraction $\rho_{2\,+}^{\star}$ of the society displays corrupt behaviors and a fraction $1-\rho_{2\,+}^{\star}$ remains honest. Finally, if both solutions of Eq. (\ref{eq:19}) are defined positive, we find two stable states $\rho_1^{\star}$ and $\rho_{2\,+}^{\star}$ separated by the unstable solution $\rho_{2\,-}^{\star}$. Note that, by breaking the symmetry including the spontaneous flow $\mu^{(0)}$, the full-corruption state disappears unless $\beta^{(m)}\rightarrow\infty\;\forall m$.
\medskip

We illustrate these results showing in Fig. \ref{fig:P4}.a-b the solutions $\rho_1^{\star}$, $\rho_{2\,+}^{\star}$ and $\rho_{2\,-}^{\star}$ as a function of $\lambda^{(1)}-\gamma^{(1)}$ for different values of $\lambda^{(2)}$ (Fig. \ref{fig:P4}.a) and $\gamma^{(2)}$  (in Fig. \ref{fig:P4}.b), fixing  $\gamma^{(2)}=0$ or  $\lambda^{(2)}=0$ respectively, and setting $\gamma^{(1)}=1$. The dashed lines describe the unstable branches given by $\rho_{2\,-}^{\star}$. In the absence of higher-order interactions, there is a continuous transition at $\lambda^{(1)}_{c,\,1}$ from the absorbent state to the active state given by $\rho_{2\,+}^{\star}$. However, Fig. \ref{fig:P4}.a shows that as $\lambda^{(2)}$ increases, the fraction of corrupt individuals in the active phase increases. Then, at a given $\lambda^{(2)}_c$, the nature of the transition changes, becoming first-order. Notably, the right limit of the bistability region is still determined by  $\lambda^{(1)}_{c,\,1}$, but the left limit depends on $\lambda^{(2)}$, i.e. $\lambda^{(1)}_{c,\,2}=\lambda^{(1)}_{c,\,2}(\lambda^{(2)})$. In contrast, when modifying $\gamma^{(2)}$ in Fig. \ref{fig:P4}.b we appreciate how, even when the transition is continuous, the critical value of $\lambda$ depends on the strength of higher-order interactions, i.e. $\lambda^{(1)}_{c,\,1}=\lambda^{(1)}_{c,\,1}(\gamma^{(2)})$. Moreover, as $\gamma^{(2)}$ increases, the transition becomes explosive when $\lambda^{(1)}_{c,\,2}=\lambda^{(1)}_{c,\,2}$.
\medskip

In Fig. \ref{fig:P4}.c we present the full phase diagram in terms of $\lambda^{(1)}-\gamma^{(1)}$ and $\bar\Delta^{(2)}=k^{(2)}\Delta^{(2)}/\mu^{(0)}$, and we identify that $\lambda^{(1)}_{c,\,1}$ and $\lambda^{(1)}_{c,\,2}$ separate three regions: fully-honest equilibrium, endemic corruption and bistability. The expressions for these boundaries are derived in the Supplementary Eqs. (S.15)-(S.19) and read as follows:
\begin{eqnarray}
    \lambda^{(1)}_{c,\,1}&=&1+\gamma^{(1)}+\gamma^{(2)},
    \label{eq:20}\\
    \lambda^{(1)}_{c,\,2}&=&2\sqrt{\lambda^{(2)}+\gamma^{(2)}}+\gamma^{(1)}-\lambda^{(2)}.
    \label{eq:21}
\end{eqnarray}
Furthermore, the onset of explosivity occurs when $\rho_{2\,+}^{\star}=\rho_{2\,-}^{\star}=0$, what is achieved (see Supplementary Eqs. (S.15)-(S.19)) provided the following condition is fulfilled: 
\begin{eqnarray}
    \lambda^{(2)}+\gamma^{(2)}=1\;.
    \label{eq:22}
\end{eqnarray}

\begin{figure}[t!]
\centering\includegraphics[width=0.975\linewidth]{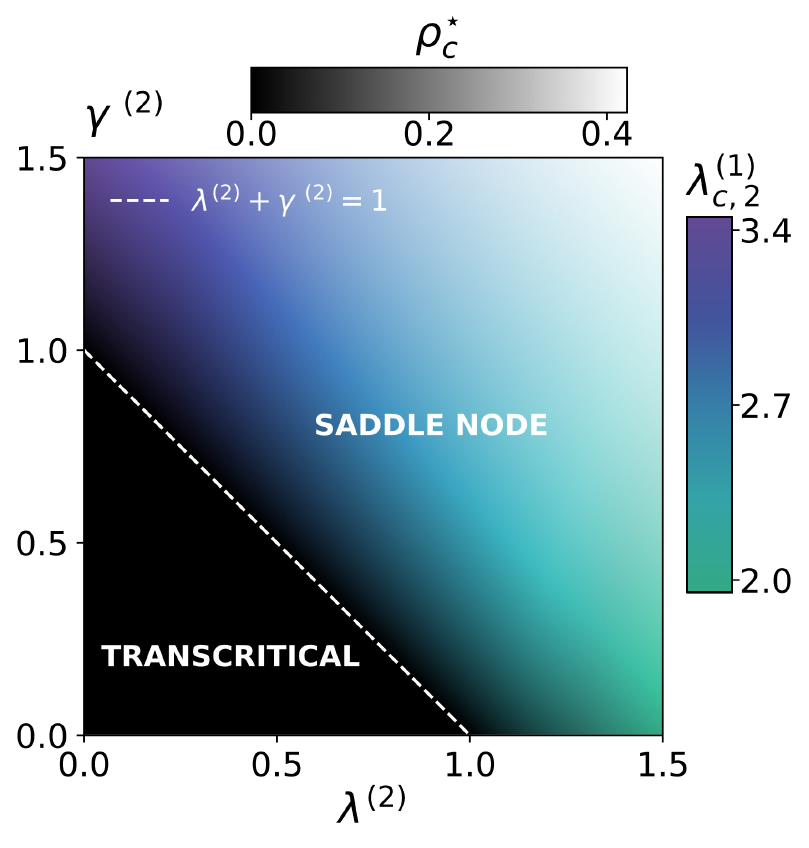}
\caption{\textbf{Bifurcation diagram of behavior adoption dynamics}. \santiago{Increasing higher-order adoption rates shifts the bifurcation type from transcritical to saddle node, and magnifies the abruptness of the transition to endemic corruption.} Color code corresponds to the critical point ($\lambda^{(1)}_{c,\,2}$, $\rho_{2\,+}^{\star}$) in terms of the higher-order adoption parameters $\lambda^{(2)}$ and $\lambda^{(2)}$. As the relevance of higher-order interaction increases the local bifurcation shifts from transcritical to saddle-node. The dashed line indicates the boundary. Note that we have set $\gamma^{(1)}=1$.}
\label{fig:P5}
\end{figure}

We condensate the findings by showing in Fig. \ref{fig:P5} the bifurcation diagram of the system. In aprticular, we represent the critical value $\lambda^{(1)}_{c,\,2}$ and the value of $\rho^{\star}_c$ after the transition (see Supplementary Eqs. (S.15)-(S.19) for the derivation), namely:
\begin{eqnarray}
\rho^{\star}_c&=&\frac{\left(\lambda^{(2)}+\gamma^{(2)}\right)-\sqrt{\lambda^{(2)}+\gamma^{(2)}}}{\left(\lambda^{(2)}+\gamma^{(2)}\right)}\;,
\end{eqnarray}
in terms of the three-body control parameters (see Supplementary Fig. 2 for a decoupled illustration of the dependencies of $\lambda^{(1)}_{c,\,2}$ and  $\rho^{\star}_c$). The black region ($\bar\Delta<1$) corresponds to the transcritical bifurcation, where the fixed points $\rho^{\star}_1$ and $\rho^{\star}_{2\,+}$ exchange their stability. Conversely, the colored area ($\bar\Delta>1$) showcases the characteristics of the critical point defining the saddle node bifurcation, where the fixed points $\rho^{\star}_{2\,+}$ and $\rho^{\star}_{2\,-}$ emerge. As $\gamma^{(2)}$ ($\lambda^{(2)
}$) increases, the critical value $\lambda^{(1)}_{c,2}$ is shifted toward larger (smaller) values. Moreover, as the higher-order coefficients increase, the abrupt jump to the endemic corruption phase widens, reaching larger values of $\rho_c^{\star}$. 
\medskip

Finally, let us note that in case recovery is not mediated by interaction ($\gamma^{(1)}=\gamma^{(2)}=0$), the phenomenology discovered by Iacopini et al. \cite{iacopini2019simplicial} is recovered. Besides, with the further assumption that $M=1$ and $\mu^{(1)}=0$, we reach the solution of the usual SIS epidemic model, with the threshold at $\beta^{(1)}k^{(1)}/\mu^{(0)}=1$ indicating the continuous transition between the absorbent state, $\rho^{\star}_1=0$, and the active state, $\rho^{\star}_{2+}=1-\mu^{(0)}/\left(\beta^{(1)}k^{(1)}\right)$.

\section{Conclusions}
\label{sec:V}

In this article, we have introduced the Higher‑Order Honesty–Corruption–Ostracism model (HO‑HCO) to investigate the impact of higher‑order interactions on the adoption of corrupt behaviors in social systems. Because higher‑order interactions naturally embody group pressure, the results are directly relevant to online social platforms, where such group effects are pervasive. Our findings reveal that the inclusion of group interactions leads to explosive transitions between predominantly honest and corrupt societies. This emergent bistability highlights the fragility of societal structures, showcasing that minor changes in the interaction dynamics can abruptly shift a population from an honest state to systemic corruption.
\medskip

Our analytical results show that this abrupt transition is absent in traditional pairwise models, where corruption takes over progressively. Furthermore, we established a connection between our results and existing models of social contagion \cite{iacopini2019simplicial}, demonstrating that when individuals exhibit a predisposition toward adopting a particular behavior, the results on the onset of explosivity align with previous findings on behavioral adoption. In fact, from a broader point of view, higher-order models of social contagion can be understood as a particular case of competition dynamics, where the acquisition of one dynamics is contact-based, and the acquisition of the other one is spontaneous.
Besides, the emergence of bistability (see Fig. \ref{fig:P3}.b) after including a synergetic mechanism in competition dynamics aligns with previous results \cite{lamata2024pathways,lucas2023simplicially,li2022competing,zhao2024susceptible}. These insights suggest that moderation policies on digital platforms should explicitly account for group‑level mechanisms if they aim to prevent abrupt shifts toward widespread misconduct.
\medskip

Moreover, the study of the case where the punishment stage is instantaneous (the limit $r\rightarrow\infty$) eases the link between this compartmental model approach and more traditional ways of modeling corruption based on game theory.  In fact, the role of honest and corrupt behaviors resembles here to the cooperator and defector strategies in the public goods game. Therefore, in the absence of higher-order interactions, the abrupt transition at $\beta^{(1)}=\mu^{(1)}$ mimics the transition to cooperation found in the public goods game \cite{perc2013evolutionary}.
\medskip

Overall, our results may have important implications for understanding the stability of social systems and the mechanisms that drive systemic corruption, as they hint that policies aimed at mitigating corruption should consider the influence of group interactions. Otherwise, interventions at the individual level may be insufficient to prevent sudden shifts toward corrupt behaviors. 
\medskip

A promising future avenue could involve incorporating the structure of pairwise \cite{vendeville2025modeling} and group-level \cite{de2024complex} friendships and enmities, since they could play an important role in the adoption of corrupt behaviors. 
Another possible avenue could involve exploring how targeted interventions in higher-order structures could stabilize societies against corruption and promote ethical behavior in complex social environments. In that regard, it would be interesting to uncover the effect of changing the microscopic \cite{lamata2025hyperedge} or temporal \cite{iacopini2024temporal,arregui2024patterns} structure of interactions.



\smallskip
\section*{Supplementary Material}
\santiago{
Supplementary material provides additional analyses supporting the results presented in the main text. In particular, we include the validation of the theoretical framework through stochastic simulations, detailed derivations of stability conditions for the fixed points, and the bifurcation diagrams characterizing the dynamical behavior of our model.}
\smallskip
\section*{Acknowledgements} 
The authors thank D. Soriano-Paños for insightful discussions on the structure of the article. S.L.O, H.P.M. and J.G.G. acknowledge support from Departamento de Industria e Innovaci\'on del Gobierno de Arag\'on y Fondo Social Europeo (FENOL group grant E36-23R) and Ministerio de Ciencia e Innovaci\'on (grant PID2023-147734NB-I00). S.L.O. and H.P.M. acknowledge financial support from Gobierno de Aragón through a doctoral fellowship. 


\section*{Conflict of interest}

The authors have no conflicts to disclose.

\section*{Author contributions}

E.B.-F., C.C.-M., V.S.-A., and G.T.-D. contributed equally to this
paper.
\medskip

\noindent\textbf{Elisa Bretón-Fuertes:} Formal analysis (equal); Investigation (equal);
Software (equal). \textbf{Clara Clemente-Marcuello:} Formal analysis
(equal); Investigation (equal); Software (equal). \textbf{Verónica Sanz-Arqué}: Formal analysis (equal); Investigation (equal); Software
(equal). \textbf{Gabriela Tomás-Delgado:} Formal analysis (equal); Investigation (equal); Software (equal). \textbf{Santiago Lamata-Otín:} Conceptualization (equal); Formal analysis (equal); Investigation (equal);
Methodology (equal); Validation (equal); Writing – original draft
(equal). \textbf{Hugo Pérez-Martinez:} Conceptualization (equal); Supervision (equal). \textbf{Jesús Gómez-Gardeñes:} Conceptualization (equal);
Methodology (equal); Project administration (equal); Supervision
(equal); Writing – review \& editing (equal).

\section*{Data Availability}

Data sharing is not applicable to this article as no new
data were created or analyzed in this study. The code is openly available in GitHub at:\\ https://github.com/santiagolaot/HO-HCO-model.

\section*{REFERENCES}

\bibliography{biblio}

\begin{thebibliography}{56}%
\makeatletter
\providecommand \@ifxundefined [1]{%
 \@ifx{#1\undefined}
}%
\providecommand \@ifnum [1]{%
 \ifnum #1\expandafter \@firstoftwo
 \else \expandafter \@secondoftwo
 \fi
}%
\providecommand \@ifx [1]{%
 \ifx #1\expandafter \@firstoftwo
 \else \expandafter \@secondoftwo
 \fi
}%
\providecommand \natexlab [1]{#1}%
\providecommand \enquote  [1]{``#1''}%
\providecommand \bibnamefont  [1]{#1}%
\providecommand \bibfnamefont [1]{#1}%
\providecommand \citenamefont [1]{#1}%
\providecommand \href@noop [0]{\@secondoftwo}%
\providecommand \href [0]{\begingroup \@sanitize@url \@href}%
\providecommand \@href[1]{\@@startlink{#1}\@@href}%
\providecommand \@@href[1]{\endgroup#1\@@endlink}%
\providecommand \@sanitize@url [0]{\catcode `\\12\catcode `\$12\catcode `\&12\catcode `\#12\catcode `\^12\catcode `\_12\catcode `\%12\relax}%
\providecommand \@@startlink[1]{}%
\providecommand \@@endlink[0]{}%
\providecommand \url  [0]{\begingroup\@sanitize@url \@url }%
\providecommand \@url [1]{\endgroup\@href {#1}{\urlprefix }}%
\providecommand \urlprefix  [0]{URL }%
\providecommand \Eprint [0]{\href }%
\providecommand \doibase [0]{http://dx.doi.org/}%
\providecommand \selectlanguage [0]{\@gobble}%
\providecommand \bibinfo  [0]{\@secondoftwo}%
\providecommand \bibfield  [0]{\@secondoftwo}%
\providecommand \translation [1]{[#1]}%
\providecommand \BibitemOpen [0]{}%
\providecommand \bibitemStop [0]{}%
\providecommand \bibitemNoStop [0]{.\EOS\space}%
\providecommand \EOS [0]{\spacefactor3000\relax}%
\providecommand \BibitemShut  [1]{\csname bibitem#1\endcsname}%
\let\auto@bib@innerbib\@empty
\bibitem [{\citenamefont {Battiston}\ \emph {et~al.}(2020)\citenamefont {Battiston}, \citenamefont {Cencetti}, \citenamefont {Iacopini}, \citenamefont {Latora}, \citenamefont {Lucas}, \citenamefont {Patania}, \citenamefont {Young},\ and\ \citenamefont {Petri}}]{battiston2020networks}%
  \BibitemOpen
  \bibfield  {author} {\bibinfo {author} {\bibfnamefont {F.}~\bibnamefont {Battiston}}, \bibinfo {author} {\bibfnamefont {G.}~\bibnamefont {Cencetti}}, \bibinfo {author} {\bibfnamefont {I.}~\bibnamefont {Iacopini}}, \bibinfo {author} {\bibfnamefont {V.}~\bibnamefont {Latora}}, \bibinfo {author} {\bibfnamefont {M.}~\bibnamefont {Lucas}}, \bibinfo {author} {\bibfnamefont {A.}~\bibnamefont {Patania}}, \bibinfo {author} {\bibfnamefont {J.-G.}\ \bibnamefont {Young}}, \ and\ \bibinfo {author} {\bibfnamefont {G.}~\bibnamefont {Petri}},\ }\bibfield  {title} {\enquote {\bibinfo {title} {Networks beyond pairwise interactions: Structure and dynamics},}\ }\href@noop {} {\bibfield  {journal} {\bibinfo  {journal} {Physics Reports}\ }\textbf {\bibinfo {volume} {874}},\ \bibinfo {pages} {1--92} (\bibinfo {year} {2020})}\BibitemShut {NoStop}%
\bibitem [{\citenamefont {Bick}\ \emph {et~al.}(2023)\citenamefont {Bick}, \citenamefont {Gross}, \citenamefont {Harrington},\ and\ \citenamefont {Schaub}}]{bick2023higher}%
  \BibitemOpen
  \bibfield  {author} {\bibinfo {author} {\bibfnamefont {C.}~\bibnamefont {Bick}}, \bibinfo {author} {\bibfnamefont {E.}~\bibnamefont {Gross}}, \bibinfo {author} {\bibfnamefont {H.~A.}\ \bibnamefont {Harrington}}, \ and\ \bibinfo {author} {\bibfnamefont {M.~T.}\ \bibnamefont {Schaub}},\ }\bibfield  {title} {\enquote {\bibinfo {title} {What are higher-order networks?}}\ }\href@noop {} {\bibfield  {journal} {\bibinfo  {journal} {SIAM Review}\ }\textbf {\bibinfo {volume} {65}},\ \bibinfo {pages} {686--731} (\bibinfo {year} {2023})}\BibitemShut {NoStop}%
\bibitem [{\citenamefont {Majhi}, \citenamefont {Perc},\ and\ \citenamefont {Ghosh}(2022)}]{majhi2022dynamics}%
  \BibitemOpen
  \bibfield  {author} {\bibinfo {author} {\bibfnamefont {S.}~\bibnamefont {Majhi}}, \bibinfo {author} {\bibfnamefont {M.}~\bibnamefont {Perc}}, \ and\ \bibinfo {author} {\bibfnamefont {D.}~\bibnamefont {Ghosh}},\ }\bibfield  {title} {\enquote {\bibinfo {title} {Dynamics on higher-order networks: A review},}\ }\href@noop {} {\bibfield  {journal} {\bibinfo  {journal} {Journal of the Royal Society Interface}\ }\textbf {\bibinfo {volume} {19}},\ \bibinfo {pages} {20220043} (\bibinfo {year} {2022})}\BibitemShut {NoStop}%
\bibitem [{\citenamefont {Battiston}\ \emph {et~al.}(2021)\citenamefont {Battiston}, \citenamefont {Amico}, \citenamefont {Barrat}, \citenamefont {Bianconi}, \citenamefont {Ferraz~de Arruda}, \citenamefont {Franceschiello}, \citenamefont {Iacopini}, \citenamefont {K{\'e}fi}, \citenamefont {Latora}, \citenamefont {Moreno} \emph {et~al.}}]{battiston2021physics}%
  \BibitemOpen
  \bibfield  {author} {\bibinfo {author} {\bibfnamefont {F.}~\bibnamefont {Battiston}}, \bibinfo {author} {\bibfnamefont {E.}~\bibnamefont {Amico}}, \bibinfo {author} {\bibfnamefont {A.}~\bibnamefont {Barrat}}, \bibinfo {author} {\bibfnamefont {G.}~\bibnamefont {Bianconi}}, \bibinfo {author} {\bibfnamefont {G.}~\bibnamefont {Ferraz~de Arruda}}, \bibinfo {author} {\bibfnamefont {B.}~\bibnamefont {Franceschiello}}, \bibinfo {author} {\bibfnamefont {I.}~\bibnamefont {Iacopini}}, \bibinfo {author} {\bibfnamefont {S.}~\bibnamefont {K{\'e}fi}}, \bibinfo {author} {\bibfnamefont {V.}~\bibnamefont {Latora}}, \bibinfo {author} {\bibfnamefont {Y.}~\bibnamefont {Moreno}},  \emph {et~al.},\ }\bibfield  {title} {\enquote {\bibinfo {title} {The physics of higher-order interactions in complex systems},}\ }\href@noop {} {\bibfield  {journal} {\bibinfo  {journal} {Nature Physics}\ }\textbf {\bibinfo {volume} {17}},\ \bibinfo {pages} {1093--1098} (\bibinfo {year} {2021})}\BibitemShut {NoStop}%
\bibitem [{\citenamefont {Iacopini}\ \emph {et~al.}(2019)\citenamefont {Iacopini}, \citenamefont {Petri}, \citenamefont {Barrat},\ and\ \citenamefont {Latora}}]{iacopini2019simplicial}%
  \BibitemOpen
  \bibfield  {author} {\bibinfo {author} {\bibfnamefont {I.}~\bibnamefont {Iacopini}}, \bibinfo {author} {\bibfnamefont {G.}~\bibnamefont {Petri}}, \bibinfo {author} {\bibfnamefont {A.}~\bibnamefont {Barrat}}, \ and\ \bibinfo {author} {\bibfnamefont {V.}~\bibnamefont {Latora}},\ }\bibfield  {title} {\enquote {\bibinfo {title} {Simplicial models of social contagion},}\ }\href@noop {} {\bibfield  {journal} {\bibinfo  {journal} {Nature communications}\ }\textbf {\bibinfo {volume} {10}},\ \bibinfo {pages} {2485} (\bibinfo {year} {2019})}\BibitemShut {NoStop}%
\bibitem [{\citenamefont {Landry}\ and\ \citenamefont {Restrepo}(2020)}]{landry2020effect}%
  \BibitemOpen
  \bibfield  {author} {\bibinfo {author} {\bibfnamefont {N.~W.}\ \bibnamefont {Landry}}\ and\ \bibinfo {author} {\bibfnamefont {J.~G.}\ \bibnamefont {Restrepo}},\ }\bibfield  {title} {\enquote {\bibinfo {title} {The effect of heterogeneity on hypergraph contagion models},}\ }\href@noop {} {\bibfield  {journal} {\bibinfo  {journal} {Chaos: An Interdisciplinary Journal of Nonlinear Science}\ }\textbf {\bibinfo {volume} {30}} (\bibinfo {year} {2020})}\BibitemShut {NoStop}%
\bibitem [{\citenamefont {St-Onge}\ \emph {et~al.}(2022)\citenamefont {St-Onge}, \citenamefont {Iacopini}, \citenamefont {Latora}, \citenamefont {Barrat}, \citenamefont {Petri}, \citenamefont {Allard},\ and\ \citenamefont {H{\'e}bert-Dufresne}}]{st2022influential}%
  \BibitemOpen
  \bibfield  {author} {\bibinfo {author} {\bibfnamefont {G.}~\bibnamefont {St-Onge}}, \bibinfo {author} {\bibfnamefont {I.}~\bibnamefont {Iacopini}}, \bibinfo {author} {\bibfnamefont {V.}~\bibnamefont {Latora}}, \bibinfo {author} {\bibfnamefont {A.}~\bibnamefont {Barrat}}, \bibinfo {author} {\bibfnamefont {G.}~\bibnamefont {Petri}}, \bibinfo {author} {\bibfnamefont {A.}~\bibnamefont {Allard}}, \ and\ \bibinfo {author} {\bibfnamefont {L.}~\bibnamefont {H{\'e}bert-Dufresne}},\ }\bibfield  {title} {\enquote {\bibinfo {title} {Influential groups for seeding and sustaining nonlinear contagion in heterogeneous hypergraphs},}\ }\href@noop {} {\bibfield  {journal} {\bibinfo  {journal} {Communications Physics}\ }\textbf {\bibinfo {volume} {5}},\ \bibinfo {pages} {25} (\bibinfo {year} {2022})}\BibitemShut {NoStop}%
\bibitem [{\citenamefont {Civilini}\ \emph {et~al.}(2024)\citenamefont {Civilini}, \citenamefont {Sadekar}, \citenamefont {Battiston}, \citenamefont {G\'omez-Garde\~nes},\ and\ \citenamefont {Latora}}]{Civilini2024explosive}%
  \BibitemOpen
  \bibfield  {author} {\bibinfo {author} {\bibfnamefont {A.}~\bibnamefont {Civilini}}, \bibinfo {author} {\bibfnamefont {O.}~\bibnamefont {Sadekar}}, \bibinfo {author} {\bibfnamefont {F.}~\bibnamefont {Battiston}}, \bibinfo {author} {\bibfnamefont {J.}~\bibnamefont {G\'omez-Garde\~nes}}, \ and\ \bibinfo {author} {\bibfnamefont {V.}~\bibnamefont {Latora}},\ }\bibfield  {title} {\enquote {\bibinfo {title} {Explosive cooperation in social dilemmas on higher-order networks},}\ }\href {\doibase 10.1103/PhysRevLett.132.167401} {\bibfield  {journal} {\bibinfo  {journal} {Phys. Rev. Lett.}\ }\textbf {\bibinfo {volume} {132}},\ \bibinfo {pages} {167401} (\bibinfo {year} {2024})}\BibitemShut {NoStop}%
\bibitem [{\citenamefont {Ferraz~de Arruda}, \citenamefont {Aleta},\ and\ \citenamefont {Moreno}(2024)}]{ferraz2024contagion}%
  \BibitemOpen
  \bibfield  {author} {\bibinfo {author} {\bibfnamefont {G.}~\bibnamefont {Ferraz~de Arruda}}, \bibinfo {author} {\bibfnamefont {A.}~\bibnamefont {Aleta}}, \ and\ \bibinfo {author} {\bibfnamefont {Y.}~\bibnamefont {Moreno}},\ }\bibfield  {title} {\enquote {\bibinfo {title} {Contagion dynamics on higher-order networks},}\ }\href@noop {} {\bibfield  {journal} {\bibinfo  {journal} {Nature Reviews Physics}\ }\textbf {\bibinfo {volume} {6}},\ \bibinfo {pages} {468--482} (\bibinfo {year} {2024})}\BibitemShut {NoStop}%
\bibitem [{\citenamefont {Malizia}\ \emph {et~al.}(2025)\citenamefont {Malizia}, \citenamefont {Lamata-Ot{\'\i}n}, \citenamefont {Frasca}, \citenamefont {Latora},\ and\ \citenamefont {G{\'o}mez-Garde{\~n}es}}]{malizia2025hyperedge}%
  \BibitemOpen
  \bibfield  {author} {\bibinfo {author} {\bibfnamefont {F.}~\bibnamefont {Malizia}}, \bibinfo {author} {\bibfnamefont {S.}~\bibnamefont {Lamata-Ot{\'\i}n}}, \bibinfo {author} {\bibfnamefont {M.}~\bibnamefont {Frasca}}, \bibinfo {author} {\bibfnamefont {V.}~\bibnamefont {Latora}}, \ and\ \bibinfo {author} {\bibfnamefont {J.}~\bibnamefont {G{\'o}mez-Garde{\~n}es}},\ }\bibfield  {title} {\enquote {\bibinfo {title} {Hyperedge overlap drives explosive transitions in systems with higher-order interactions},}\ }\href@noop {} {\bibfield  {journal} {\bibinfo  {journal} {Nature communications}\ }\textbf {\bibinfo {volume} {16}},\ \bibinfo {pages} {555} (\bibinfo {year} {2025})}\BibitemShut {NoStop}%
\bibitem [{\citenamefont {Alvarez-Rodriguez}\ \emph {et~al.}(2021)\citenamefont {Alvarez-Rodriguez}, \citenamefont {Battiston}, \citenamefont {de~Arruda}, \citenamefont {Moreno}, \citenamefont {Perc},\ and\ \citenamefont {Latora}}]{alvarez2021evolutionary}%
  \BibitemOpen
  \bibfield  {author} {\bibinfo {author} {\bibfnamefont {U.}~\bibnamefont {Alvarez-Rodriguez}}, \bibinfo {author} {\bibfnamefont {F.}~\bibnamefont {Battiston}}, \bibinfo {author} {\bibfnamefont {G.~F.}\ \bibnamefont {de~Arruda}}, \bibinfo {author} {\bibfnamefont {Y.}~\bibnamefont {Moreno}}, \bibinfo {author} {\bibfnamefont {M.}~\bibnamefont {Perc}}, \ and\ \bibinfo {author} {\bibfnamefont {V.}~\bibnamefont {Latora}},\ }\bibfield  {title} {\enquote {\bibinfo {title} {Evolutionary dynamics of higher-order interactions in social networks},}\ }\href@noop {} {\bibfield  {journal} {\bibinfo  {journal} {Nature Human Behaviour}\ }\textbf {\bibinfo {volume} {5}},\ \bibinfo {pages} {586--595} (\bibinfo {year} {2021})}\BibitemShut {NoStop}%
\bibitem [{\citenamefont {Dong}\ \emph {et~al.}(2025)\citenamefont {Dong}, \citenamefont {Huo}, \citenamefont {Perc},\ and\ \citenamefont {Boccaletti}}]{dong2025adaptive}%
  \BibitemOpen
  \bibfield  {author} {\bibinfo {author} {\bibfnamefont {Y.}~\bibnamefont {Dong}}, \bibinfo {author} {\bibfnamefont {L.}~\bibnamefont {Huo}}, \bibinfo {author} {\bibfnamefont {M.}~\bibnamefont {Perc}}, \ and\ \bibinfo {author} {\bibfnamefont {S.}~\bibnamefont {Boccaletti}},\ }\bibfield  {title} {\enquote {\bibinfo {title} {Adaptive rumor propagation and activity contagion in higher-order networks},}\ }\href@noop {} {\bibfield  {journal} {\bibinfo  {journal} {Communications Physics}\ }\textbf {\bibinfo {volume} {8}},\ \bibinfo {pages} {261} (\bibinfo {year} {2025})}\BibitemShut {NoStop}%
\bibitem [{\citenamefont {Vosoughi}, \citenamefont {Roy},\ and\ \citenamefont {Aral}(2018)}]{vosoughi2018spread}%
  \BibitemOpen
  \bibfield  {author} {\bibinfo {author} {\bibfnamefont {S.}~\bibnamefont {Vosoughi}}, \bibinfo {author} {\bibfnamefont {D.}~\bibnamefont {Roy}}, \ and\ \bibinfo {author} {\bibfnamefont {S.}~\bibnamefont {Aral}},\ }\bibfield  {title} {\enquote {\bibinfo {title} {The spread of true and false news online},}\ }\href@noop {} {\bibfield  {journal} {\bibinfo  {journal} {science}\ }\textbf {\bibinfo {volume} {359}},\ \bibinfo {pages} {1146--1151} (\bibinfo {year} {2018})}\BibitemShut {NoStop}%
\bibitem [{\citenamefont {Bakshy}\ \emph {et~al.}(2012)\citenamefont {Bakshy}, \citenamefont {Rosenn}, \citenamefont {Marlow},\ and\ \citenamefont {Adamic}}]{bakshy2012role}%
  \BibitemOpen
  \bibfield  {author} {\bibinfo {author} {\bibfnamefont {E.}~\bibnamefont {Bakshy}}, \bibinfo {author} {\bibfnamefont {I.}~\bibnamefont {Rosenn}}, \bibinfo {author} {\bibfnamefont {C.}~\bibnamefont {Marlow}}, \ and\ \bibinfo {author} {\bibfnamefont {L.}~\bibnamefont {Adamic}},\ }\bibfield  {title} {\enquote {\bibinfo {title} {The role of social networks in information diffusion},}\ }in\ \href@noop {} {\emph {\bibinfo {booktitle} {Proceedings of the 21st international conference on World Wide Web}}}\ (\bibinfo {year} {2012})\ pp.\ \bibinfo {pages} {519--528}\BibitemShut {NoStop}%
\bibitem [{\citenamefont {Aral}, \citenamefont {Muchnik},\ and\ \citenamefont {Sundararajan}(2009)}]{aral2009distinguishing}%
  \BibitemOpen
  \bibfield  {author} {\bibinfo {author} {\bibfnamefont {S.}~\bibnamefont {Aral}}, \bibinfo {author} {\bibfnamefont {L.}~\bibnamefont {Muchnik}}, \ and\ \bibinfo {author} {\bibfnamefont {A.}~\bibnamefont {Sundararajan}},\ }\bibfield  {title} {\enquote {\bibinfo {title} {Distinguishing influence-based contagion from homophily-driven diffusion in dynamic networks},}\ }\href@noop {} {\bibfield  {journal} {\bibinfo  {journal} {Proceedings of the National Academy of Sciences}\ }\textbf {\bibinfo {volume} {106}},\ \bibinfo {pages} {21544--21549} (\bibinfo {year} {2009})}\BibitemShut {NoStop}%
\bibitem [{\citenamefont {Castellano}, \citenamefont {Fortunato},\ and\ \citenamefont {Loreto}(2009)}]{castellano2009statistical}%
  \BibitemOpen
  \bibfield  {author} {\bibinfo {author} {\bibfnamefont {C.}~\bibnamefont {Castellano}}, \bibinfo {author} {\bibfnamefont {S.}~\bibnamefont {Fortunato}}, \ and\ \bibinfo {author} {\bibfnamefont {V.}~\bibnamefont {Loreto}},\ }\bibfield  {title} {\enquote {\bibinfo {title} {Statistical physics of social dynamics},}\ }\href@noop {} {\bibfield  {journal} {\bibinfo  {journal} {Reviews of modern physics}\ }\textbf {\bibinfo {volume} {81}},\ \bibinfo {pages} {591--646} (\bibinfo {year} {2009})}\BibitemShut {NoStop}%
\bibitem [{\citenamefont {Axelrod}(1997)}]{axelrod1997dissemination}%
  \BibitemOpen
  \bibfield  {author} {\bibinfo {author} {\bibfnamefont {R.}~\bibnamefont {Axelrod}},\ }\bibfield  {title} {\enquote {\bibinfo {title} {The dissemination of culture: A model with local convergence and global polarization},}\ }\href@noop {} {\bibfield  {journal} {\bibinfo  {journal} {Journal of conflict resolution}\ }\textbf {\bibinfo {volume} {41}},\ \bibinfo {pages} {203--226} (\bibinfo {year} {1997})}\BibitemShut {NoStop}%
\bibitem [{\citenamefont {Granovetter}(1978)}]{granovetter1978threshold}%
  \BibitemOpen
  \bibfield  {author} {\bibinfo {author} {\bibfnamefont {M.}~\bibnamefont {Granovetter}},\ }\bibfield  {title} {\enquote {\bibinfo {title} {Threshold models of collective behavior},}\ }\href@noop {} {\bibfield  {journal} {\bibinfo  {journal} {American journal of sociology}\ }\textbf {\bibinfo {volume} {83}},\ \bibinfo {pages} {1420--1443} (\bibinfo {year} {1978})}\BibitemShut {NoStop}%
\bibitem [{\citenamefont {Centola}(2010)}]{centola2010spread}%
  \BibitemOpen
  \bibfield  {author} {\bibinfo {author} {\bibfnamefont {D.}~\bibnamefont {Centola}},\ }\bibfield  {title} {\enquote {\bibinfo {title} {The spread of behavior in an online social network experiment},}\ }\href@noop {} {\bibfield  {journal} {\bibinfo  {journal} {science}\ }\textbf {\bibinfo {volume} {329}},\ \bibinfo {pages} {1194--1197} (\bibinfo {year} {2010})}\BibitemShut {NoStop}%
\bibitem [{\citenamefont {Daley}\ and\ \citenamefont {Kendall}(1964)}]{daley1964epidemics}%
  \BibitemOpen
  \bibfield  {author} {\bibinfo {author} {\bibfnamefont {D.~J.}\ \bibnamefont {Daley}}\ and\ \bibinfo {author} {\bibfnamefont {D.~G.}\ \bibnamefont {Kendall}},\ }\bibfield  {title} {\enquote {\bibinfo {title} {Epidemics and rumours},}\ }\href@noop {} {\bibfield  {journal} {\bibinfo  {journal} {Nature}\ }\textbf {\bibinfo {volume} {204}},\ \bibinfo {pages} {1118--1118} (\bibinfo {year} {1964})}\BibitemShut {NoStop}%
\bibitem [{\citenamefont {Maki}(1973)}]{maki1973mathematical}%
  \BibitemOpen
  \bibfield  {author} {\bibinfo {author} {\bibfnamefont {D.}~\bibnamefont {Maki}},\ }\href@noop {} {\enquote {\bibinfo {title} {Mathematical models and applications, with emphasis on the social, life, and management sciences},}\ } (\bibinfo {year} {1973})\BibitemShut {NoStop}%
\bibitem [{\citenamefont {Starnini}\ \emph {et~al.}(2025)\citenamefont {Starnini}, \citenamefont {Baumann}, \citenamefont {Galla}, \citenamefont {Garcia}, \citenamefont {I{\~n}iguez}, \citenamefont {Karsai}, \citenamefont {Lorenz},\ and\ \citenamefont {Sznajd-Weron}}]{starnini2025opinion}%
  \BibitemOpen
  \bibfield  {author} {\bibinfo {author} {\bibfnamefont {M.}~\bibnamefont {Starnini}}, \bibinfo {author} {\bibfnamefont {F.}~\bibnamefont {Baumann}}, \bibinfo {author} {\bibfnamefont {T.}~\bibnamefont {Galla}}, \bibinfo {author} {\bibfnamefont {D.}~\bibnamefont {Garcia}}, \bibinfo {author} {\bibfnamefont {G.}~\bibnamefont {I{\~n}iguez}}, \bibinfo {author} {\bibfnamefont {M.}~\bibnamefont {Karsai}}, \bibinfo {author} {\bibfnamefont {J.}~\bibnamefont {Lorenz}}, \ and\ \bibinfo {author} {\bibfnamefont {K.}~\bibnamefont {Sznajd-Weron}},\ }\bibfield  {title} {\enquote {\bibinfo {title} {Opinion dynamics: Statistical physics and beyond},}\ }\href@noop {} {\bibfield  {journal} {\bibinfo  {journal} {arXiv preprint arXiv:2507.11521}\ } (\bibinfo {year} {2025})}\BibitemShut {NoStop}%
\bibitem [{\citenamefont {P{\'e}rez-Mart{\'\i}nez}\ \emph {et~al.}(2025)\citenamefont {P{\'e}rez-Mart{\'\i}nez}, \citenamefont {Lamata-Ot{\'\i}n}, \citenamefont {Malizia}, \citenamefont {Flor{\'\i}a}, \citenamefont {G{\'o}mez-Garde{\~n}es},\ and\ \citenamefont {Soriano-Pa{\~n}os}}]{perez2025social}%
  \BibitemOpen
  \bibfield  {author} {\bibinfo {author} {\bibfnamefont {H.}~\bibnamefont {P{\'e}rez-Mart{\'\i}nez}}, \bibinfo {author} {\bibfnamefont {S.}~\bibnamefont {Lamata-Ot{\'\i}n}}, \bibinfo {author} {\bibfnamefont {F.}~\bibnamefont {Malizia}}, \bibinfo {author} {\bibfnamefont {L.~M.}\ \bibnamefont {Flor{\'\i}a}}, \bibinfo {author} {\bibfnamefont {J.}~\bibnamefont {G{\'o}mez-Garde{\~n}es}}, \ and\ \bibinfo {author} {\bibfnamefont {D.}~\bibnamefont {Soriano-Pa{\~n}os}},\ }\bibfield  {title} {\enquote {\bibinfo {title} {Social polarization promoted by sparse higher-order interactions},}\ }\href@noop {} {\bibfield  {journal} {\bibinfo  {journal} {arXiv preprint arXiv:2507.12325}\ } (\bibinfo {year} {2025})}\BibitemShut {NoStop}%
\bibitem [{\citenamefont {Axelrod}\ and\ \citenamefont {Hamilton}(1981)}]{axelrod1981evolution}%
  \BibitemOpen
  \bibfield  {author} {\bibinfo {author} {\bibfnamefont {R.}~\bibnamefont {Axelrod}}\ and\ \bibinfo {author} {\bibfnamefont {W.~D.}\ \bibnamefont {Hamilton}},\ }\bibfield  {title} {\enquote {\bibinfo {title} {The evolution of cooperation},}\ }\href@noop {} {\bibfield  {journal} {\bibinfo  {journal} {science}\ }\textbf {\bibinfo {volume} {211}},\ \bibinfo {pages} {1390--1396} (\bibinfo {year} {1981})}\BibitemShut {NoStop}%
\bibitem [{\citenamefont {{United Nations}}(2018)}]{un_corruption_costs_2018}%
  \BibitemOpen
  \bibfield  {author} {\bibinfo {author} {\bibnamefont {{United Nations}}},\ }\href {https://press.un.org/en/2018/sc13493.doc.htm} {\enquote {\bibinfo {title} {Global cost of corruption at least 5 per cent of world gross domestic product, secretary-general tells security council, citing world economic forum data},}\ } (\bibinfo {year} {2018}),\ \bibinfo {note} {accessed: 2024-12-13}\BibitemShut {NoStop}%
\bibitem [{\citenamefont {Ipsos}(2024)}]{ipsos2024}%
  \BibitemOpen
  \bibfield  {author} {\bibinfo {author} {\bibnamefont {Ipsos}},\ }\href {https://www.ipsos.com/es-es/desempleo-salud-e-inmigracion-los-tres-asuntos-que-mas-preocupan-la-poblacion-espanola} {\enquote {\bibinfo {title} {What worries the world},}\ } (\bibinfo {year} {2024}),\ \bibinfo {note} {accessed: 2024-11-29}\BibitemShut {NoStop}%
\bibitem [{\citenamefont {Martins}\ \emph {et~al.}(2022)\citenamefont {Martins}, \citenamefont {da~Cunha}, \citenamefont {Hanley}, \citenamefont {Gon{\c{c}}alves}, \citenamefont {Perc},\ and\ \citenamefont {Ribeiro}}]{martins2022universality}%
  \BibitemOpen
  \bibfield  {author} {\bibinfo {author} {\bibfnamefont {A.~F.}\ \bibnamefont {Martins}}, \bibinfo {author} {\bibfnamefont {B.~R.}\ \bibnamefont {da~Cunha}}, \bibinfo {author} {\bibfnamefont {Q.~S.}\ \bibnamefont {Hanley}}, \bibinfo {author} {\bibfnamefont {S.}~\bibnamefont {Gon{\c{c}}alves}}, \bibinfo {author} {\bibfnamefont {M.}~\bibnamefont {Perc}}, \ and\ \bibinfo {author} {\bibfnamefont {H.~V.}\ \bibnamefont {Ribeiro}},\ }\bibfield  {title} {\enquote {\bibinfo {title} {Universality of political corruption networks},}\ }\href@noop {} {\bibfield  {journal} {\bibinfo  {journal} {Scientific Reports}\ }\textbf {\bibinfo {volume} {12}},\ \bibinfo {pages} {6858} (\bibinfo {year} {2022})}\BibitemShut {NoStop}%
\bibitem [{\citenamefont {Pessa}\ \emph {et~al.}(2025)\citenamefont {Pessa}, \citenamefont {Martins}, \citenamefont {Prates}, \citenamefont {Gon{\c{c}}alves}, \citenamefont {Masoller}, \citenamefont {Perc},\ and\ \citenamefont {Ribeiro}}]{pessa2025structural}%
  \BibitemOpen
  \bibfield  {author} {\bibinfo {author} {\bibfnamefont {A.~A.}\ \bibnamefont {Pessa}}, \bibinfo {author} {\bibfnamefont {A.~F.}\ \bibnamefont {Martins}}, \bibinfo {author} {\bibfnamefont {M.~V.}\ \bibnamefont {Prates}}, \bibinfo {author} {\bibfnamefont {S.}~\bibnamefont {Gon{\c{c}}alves}}, \bibinfo {author} {\bibfnamefont {C.}~\bibnamefont {Masoller}}, \bibinfo {author} {\bibfnamefont {M.}~\bibnamefont {Perc}}, \ and\ \bibinfo {author} {\bibfnamefont {H.~V.}\ \bibnamefont {Ribeiro}},\ }\bibfield  {title} {\enquote {\bibinfo {title} {Structural roles and gender disparities in corruption networks},}\ }\href@noop {} {\bibfield  {journal} {\bibinfo  {journal} {Scientific reports}\ }\textbf {\bibinfo {volume} {15}},\ \bibinfo {pages} {14201} (\bibinfo {year} {2025})}\BibitemShut {NoStop}%
\bibitem [{\citenamefont {Lee}\ \emph {et~al.}(2019)\citenamefont {Lee}, \citenamefont {Iwasa}, \citenamefont {Dieckmann},\ and\ \citenamefont {Sigmund}}]{lee2019social}%
  \BibitemOpen
  \bibfield  {author} {\bibinfo {author} {\bibfnamefont {J.-H.}\ \bibnamefont {Lee}}, \bibinfo {author} {\bibfnamefont {Y.}~\bibnamefont {Iwasa}}, \bibinfo {author} {\bibfnamefont {U.}~\bibnamefont {Dieckmann}}, \ and\ \bibinfo {author} {\bibfnamefont {K.}~\bibnamefont {Sigmund}},\ }\bibfield  {title} {\enquote {\bibinfo {title} {Social evolution leads to persistent corruption},}\ }\href@noop {} {\bibfield  {journal} {\bibinfo  {journal} {Proceedings of the National Academy of Sciences}\ }\textbf {\bibinfo {volume} {116}},\ \bibinfo {pages} {13276--13281} (\bibinfo {year} {2019})}\BibitemShut {NoStop}%
\bibitem [{\citenamefont {Kolokoltsov}(2012)}]{kolokoltsov2012nonlinear}%
  \BibitemOpen
  \bibfield  {author} {\bibinfo {author} {\bibfnamefont {V.~N.}\ \bibnamefont {Kolokoltsov}},\ }\bibfield  {title} {\enquote {\bibinfo {title} {Nonlinear markov games on a finite state space (mean-field and binary interactions)},}\ }\href@noop {} {\bibfield  {journal} {\bibinfo  {journal} {International Journal of Statistics and Probability}\ }\textbf {\bibinfo {volume} {1}} (\bibinfo {year} {2012})}\BibitemShut {NoStop}%
\bibitem [{\citenamefont {Kolokoltsov}\ and\ \citenamefont {Malafeyev}(2017)}]{kolokoltsov2017mean}%
  \BibitemOpen
  \bibfield  {author} {\bibinfo {author} {\bibfnamefont {V.~N.}\ \bibnamefont {Kolokoltsov}}\ and\ \bibinfo {author} {\bibfnamefont {O.~A.}\ \bibnamefont {Malafeyev}},\ }\bibfield  {title} {\enquote {\bibinfo {title} {Mean-field-game model of corruption},}\ }\href@noop {} {\bibfield  {journal} {\bibinfo  {journal} {Dynamic Games and Applications}\ }\textbf {\bibinfo {volume} {7}},\ \bibinfo {pages} {34--47} (\bibinfo {year} {2017})}\BibitemShut {NoStop}%
\bibitem [{\citenamefont {Lee}\ \emph {et~al.}(2015)\citenamefont {Lee}, \citenamefont {Sigmund}, \citenamefont {Dieckmann},\ and\ \citenamefont {Iwasa}}]{lee2015games}%
  \BibitemOpen
  \bibfield  {author} {\bibinfo {author} {\bibfnamefont {J.-H.}\ \bibnamefont {Lee}}, \bibinfo {author} {\bibfnamefont {K.}~\bibnamefont {Sigmund}}, \bibinfo {author} {\bibfnamefont {U.}~\bibnamefont {Dieckmann}}, \ and\ \bibinfo {author} {\bibfnamefont {Y.}~\bibnamefont {Iwasa}},\ }\bibfield  {title} {\enquote {\bibinfo {title} {Games of corruption: How to suppress illegal logging},}\ }\href@noop {} {\bibfield  {journal} {\bibinfo  {journal} {Journal of Theoretical Biology}\ }\textbf {\bibinfo {volume} {367}},\ \bibinfo {pages} {1--13} (\bibinfo {year} {2015})}\BibitemShut {NoStop}%
\bibitem [{\citenamefont {Lee}, \citenamefont {Jusup},\ and\ \citenamefont {Iwasa}(2017)}]{lee2017games}%
  \BibitemOpen
  \bibfield  {author} {\bibinfo {author} {\bibfnamefont {J.-H.}\ \bibnamefont {Lee}}, \bibinfo {author} {\bibfnamefont {M.}~\bibnamefont {Jusup}}, \ and\ \bibinfo {author} {\bibfnamefont {Y.}~\bibnamefont {Iwasa}},\ }\bibfield  {title} {\enquote {\bibinfo {title} {Games of corruption in preventing the overuse of common-pool resources},}\ }\href@noop {} {\bibfield  {journal} {\bibinfo  {journal} {Journal of theoretical biology}\ }\textbf {\bibinfo {volume} {428}},\ \bibinfo {pages} {76--86} (\bibinfo {year} {2017})}\BibitemShut {NoStop}%
\bibitem [{\citenamefont {Verma}\ and\ \citenamefont {Sengupta}(2015)}]{verma2015bribe}%
  \BibitemOpen
  \bibfield  {author} {\bibinfo {author} {\bibfnamefont {P.}~\bibnamefont {Verma}}\ and\ \bibinfo {author} {\bibfnamefont {S.}~\bibnamefont {Sengupta}},\ }\bibfield  {title} {\enquote {\bibinfo {title} {Bribe and punishment: An evolutionary game-theoretic analysis of bribery},}\ }\href@noop {} {\bibfield  {journal} {\bibinfo  {journal} {PLoS One}\ }\textbf {\bibinfo {volume} {10}},\ \bibinfo {pages} {e0133441} (\bibinfo {year} {2015})}\BibitemShut {NoStop}%
\bibitem [{\citenamefont {Verma}, \citenamefont {Nandi},\ and\ \citenamefont {Sengupta}(2017)}]{verma2017bribery}%
  \BibitemOpen
  \bibfield  {author} {\bibinfo {author} {\bibfnamefont {P.}~\bibnamefont {Verma}}, \bibinfo {author} {\bibfnamefont {A.~K.}\ \bibnamefont {Nandi}}, \ and\ \bibinfo {author} {\bibfnamefont {S.}~\bibnamefont {Sengupta}},\ }\bibfield  {title} {\enquote {\bibinfo {title} {Bribery games on inter-dependent regular networks},}\ }\href@noop {} {\bibfield  {journal} {\bibinfo  {journal} {Scientific Reports}\ }\textbf {\bibinfo {volume} {7}},\ \bibinfo {pages} {42735} (\bibinfo {year} {2017})}\BibitemShut {NoStop}%
\bibitem [{\citenamefont {Verma}, \citenamefont {Nandi},\ and\ \citenamefont {Sengupta}(2018)}]{verma2018bribery}%
  \BibitemOpen
  \bibfield  {author} {\bibinfo {author} {\bibfnamefont {P.}~\bibnamefont {Verma}}, \bibinfo {author} {\bibfnamefont {A.~K.}\ \bibnamefont {Nandi}}, \ and\ \bibinfo {author} {\bibfnamefont {S.}~\bibnamefont {Sengupta}},\ }\bibfield  {title} {\enquote {\bibinfo {title} {Bribery games on interdependent complex networks},}\ }\href@noop {} {\bibfield  {journal} {\bibinfo  {journal} {Journal of Theoretical Biology}\ }\textbf {\bibinfo {volume} {450}},\ \bibinfo {pages} {43--52} (\bibinfo {year} {2018})}\BibitemShut {NoStop}%
\bibitem [{\citenamefont {Von~Neumann}\ and\ \citenamefont {Morgenstern}(2007)}]{von2007theory}%
  \BibitemOpen
  \bibfield  {author} {\bibinfo {author} {\bibfnamefont {J.}~\bibnamefont {Von~Neumann}}\ and\ \bibinfo {author} {\bibfnamefont {O.}~\bibnamefont {Morgenstern}},\ }\bibfield  {title} {\enquote {\bibinfo {title} {Theory of games and economic behavior: 60th anniversary commemorative edition},}\ }in\ \href@noop {} {\emph {\bibinfo {booktitle} {Theory of games and economic behavior}}}\ (\bibinfo  {publisher} {Princeton university press},\ \bibinfo {year} {2007})\BibitemShut {NoStop}%
\bibitem [{\citenamefont {Lu}\ \emph {et~al.}(2020)\citenamefont {Lu}, \citenamefont {Bauza}, \citenamefont {Soriano-Pa{\~n}os}, \citenamefont {G{\'o}mez-Garde{\~n}es},\ and\ \citenamefont {Flor{\'\i}a}}]{lu2020norm}%
  \BibitemOpen
  \bibfield  {author} {\bibinfo {author} {\bibfnamefont {D.}~\bibnamefont {Lu}}, \bibinfo {author} {\bibfnamefont {F.}~\bibnamefont {Bauza}}, \bibinfo {author} {\bibfnamefont {D.}~\bibnamefont {Soriano-Pa{\~n}os}}, \bibinfo {author} {\bibfnamefont {J.}~\bibnamefont {G{\'o}mez-Garde{\~n}es}}, \ and\ \bibinfo {author} {\bibfnamefont {L.}~\bibnamefont {Flor{\'\i}a}},\ }\bibfield  {title} {\enquote {\bibinfo {title} {Norm violation versus punishment risk in a social model of corruption},}\ }\href@noop {} {\bibfield  {journal} {\bibinfo  {journal} {Physical Review E}\ }\textbf {\bibinfo {volume} {101}},\ \bibinfo {pages} {022306} (\bibinfo {year} {2020})}\BibitemShut {NoStop}%
\bibitem [{\citenamefont {Bauz{\'a}}\ \emph {et~al.}(2020)\citenamefont {Bauz{\'a}}, \citenamefont {Soriano-Pa{\~n}os}, \citenamefont {G{\'o}mez-Garde{\~n}es},\ and\ \citenamefont {Flor{\'\i}a}}]{bauza2020fear}%
  \BibitemOpen
  \bibfield  {author} {\bibinfo {author} {\bibfnamefont {F.}~\bibnamefont {Bauz{\'a}}}, \bibinfo {author} {\bibfnamefont {D.}~\bibnamefont {Soriano-Pa{\~n}os}}, \bibinfo {author} {\bibfnamefont {J.}~\bibnamefont {G{\'o}mez-Garde{\~n}es}}, \ and\ \bibinfo {author} {\bibfnamefont {L.}~\bibnamefont {Flor{\'\i}a}},\ }\bibfield  {title} {\enquote {\bibinfo {title} {Fear induced explosive transitions in the dynamics of corruption},}\ }\href@noop {} {\bibfield  {journal} {\bibinfo  {journal} {Chaos: An Interdisciplinary Journal of Nonlinear Science}\ }\textbf {\bibinfo {volume} {30}} (\bibinfo {year} {2020})}\BibitemShut {NoStop}%
\bibitem [{\citenamefont {P{\'e}rez-Mart{\'\i}nez}\ \emph {et~al.}(2022)\citenamefont {P{\'e}rez-Mart{\'\i}nez}, \citenamefont {Bauz{\'a}}, \citenamefont {Soriano-Pa{\~n}os}, \citenamefont {G{\'o}mez-Garde{\~n}es},\ and\ \citenamefont {Flor{\'\i}a}}]{perez2022emergence}%
  \BibitemOpen
  \bibfield  {author} {\bibinfo {author} {\bibfnamefont {H.}~\bibnamefont {P{\'e}rez-Mart{\'\i}nez}}, \bibinfo {author} {\bibfnamefont {F.}~\bibnamefont {Bauz{\'a}}}, \bibinfo {author} {\bibfnamefont {D.}~\bibnamefont {Soriano-Pa{\~n}os}}, \bibinfo {author} {\bibfnamefont {J.}~\bibnamefont {G{\'o}mez-Garde{\~n}es}}, \ and\ \bibinfo {author} {\bibfnamefont {L.}~\bibnamefont {Flor{\'\i}a}},\ }\bibfield  {title} {\enquote {\bibinfo {title} {Emergence, survival, and segregation of competing gangs},}\ }\href@noop {} {\bibfield  {journal} {\bibinfo  {journal} {Chaos: An Interdisciplinary Journal of Nonlinear Science}\ }\textbf {\bibinfo {volume} {32}} (\bibinfo {year} {2022})}\BibitemShut {NoStop}%
\bibitem [{\citenamefont {Lu}\ and\ \citenamefont {Floria}(2023)}]{lu2023dynamics}%
  \BibitemOpen
  \bibfield  {author} {\bibinfo {author} {\bibfnamefont {D.}~\bibnamefont {Lu}}\ and\ \bibinfo {author} {\bibfnamefont {L.~M.}\ \bibnamefont {Floria}},\ }\bibfield  {title} {\enquote {\bibinfo {title} {Dynamics of corruption on correlated multiplex networks with overlap},}\ }\href@noop {} {\bibfield  {journal} {\bibinfo  {journal} {Chaos, Solitons \& Fractals}\ }\textbf {\bibinfo {volume} {171}},\ \bibinfo {pages} {113432} (\bibinfo {year} {2023})}\BibitemShut {NoStop}%
\bibitem [{\citenamefont {Kermack}\ and\ \citenamefont {McKendrick}(1927)}]{kermack1927contribution}%
  \BibitemOpen
  \bibfield  {author} {\bibinfo {author} {\bibfnamefont {W.~O.}\ \bibnamefont {Kermack}}\ and\ \bibinfo {author} {\bibfnamefont {A.~G.}\ \bibnamefont {McKendrick}},\ }\bibfield  {title} {\enquote {\bibinfo {title} {A contribution to the mathematical theory of epidemics},}\ }\href@noop {} {\bibfield  {journal} {\bibinfo  {journal} {Proceedings of the royal society of london. Series A, Containing papers of a mathematical and physical character}\ }\textbf {\bibinfo {volume} {115}},\ \bibinfo {pages} {700--721} (\bibinfo {year} {1927})}\BibitemShut {NoStop}%
\bibitem [{\citenamefont {Barrat}, \citenamefont {Barthelemy},\ and\ \citenamefont {Vespignani}(2008)}]{barrat2008dynamical}%
  \BibitemOpen
  \bibfield  {author} {\bibinfo {author} {\bibfnamefont {A.}~\bibnamefont {Barrat}}, \bibinfo {author} {\bibfnamefont {M.}~\bibnamefont {Barthelemy}}, \ and\ \bibinfo {author} {\bibfnamefont {A.}~\bibnamefont {Vespignani}},\ }\href@noop {} {\emph {\bibinfo {title} {Dynamical processes on complex networks}}}\ (\bibinfo  {publisher} {Cambridge university press},\ \bibinfo {year} {2008})\BibitemShut {NoStop}%
\bibitem [{\citenamefont {Boccara}\ and\ \citenamefont {Boccara}(2010)}]{boccara2010modeling}%
  \BibitemOpen
  \bibfield  {author} {\bibinfo {author} {\bibfnamefont {N.}~\bibnamefont {Boccara}}\ and\ \bibinfo {author} {\bibfnamefont {N.}~\bibnamefont {Boccara}},\ }\href@noop {} {\emph {\bibinfo {title} {Modeling complex systems}}},\ Vol.~\bibinfo {volume} {1}\ (\bibinfo  {publisher} {Springer},\ \bibinfo {year} {2010})\BibitemShut {NoStop}%
\bibitem [{\citenamefont {Granell}\ \emph {et~al.}(2024)\citenamefont {Granell}, \citenamefont {G{\'o}mez}, \citenamefont {G{\'o}mez-Garde{\~n}es},\ and\ \citenamefont {Arenas}}]{granell2024probabilistic}%
  \BibitemOpen
  \bibfield  {author} {\bibinfo {author} {\bibfnamefont {C.}~\bibnamefont {Granell}}, \bibinfo {author} {\bibfnamefont {S.}~\bibnamefont {G{\'o}mez}}, \bibinfo {author} {\bibfnamefont {J.}~\bibnamefont {G{\'o}mez-Garde{\~n}es}}, \ and\ \bibinfo {author} {\bibfnamefont {A.}~\bibnamefont {Arenas}},\ }\bibfield  {title} {\enquote {\bibinfo {title} {Probabilistic discrete-time models for spreading processes in complex networks: A review},}\ }\href@noop {} {\bibfield  {journal} {\bibinfo  {journal} {Annalen der Physik}\ }\textbf {\bibinfo {volume} {536}},\ \bibinfo {pages} {2400078} (\bibinfo {year} {2024})}\BibitemShut {NoStop}%
\bibitem [{\citenamefont {Lamata-Ot{\'\i}n}, \citenamefont {G{\'o}mez-Garde{\~n}es},\ and\ \citenamefont {Soriano-Pa{\~n}os}(2024)}]{lamata2024pathways}%
  \BibitemOpen
  \bibfield  {author} {\bibinfo {author} {\bibfnamefont {S.}~\bibnamefont {Lamata-Ot{\'\i}n}}, \bibinfo {author} {\bibfnamefont {J.}~\bibnamefont {G{\'o}mez-Garde{\~n}es}}, \ and\ \bibinfo {author} {\bibfnamefont {D.}~\bibnamefont {Soriano-Pa{\~n}os}},\ }\bibfield  {title} {\enquote {\bibinfo {title} {Pathways to discontinuous transitions in interacting contagion dynamics},}\ }\href {\doibase 10.1088/2632-072X/ad269b} {\bibfield  {journal} {\bibinfo  {journal} {Journal of Physics: Complexity}\ }\textbf {\bibinfo {volume} {5}},\ \bibinfo {pages} {015015} (\bibinfo {year} {2024})}\BibitemShut {NoStop}%
\bibitem [{\citenamefont {Lucas}\ \emph {et~al.}(2023)\citenamefont {Lucas}, \citenamefont {Iacopini}, \citenamefont {Robiglio}, \citenamefont {Barrat},\ and\ \citenamefont {Petri}}]{lucas2023simplicially}%
  \BibitemOpen
  \bibfield  {author} {\bibinfo {author} {\bibfnamefont {M.}~\bibnamefont {Lucas}}, \bibinfo {author} {\bibfnamefont {I.}~\bibnamefont {Iacopini}}, \bibinfo {author} {\bibfnamefont {T.}~\bibnamefont {Robiglio}}, \bibinfo {author} {\bibfnamefont {A.}~\bibnamefont {Barrat}}, \ and\ \bibinfo {author} {\bibfnamefont {G.}~\bibnamefont {Petri}},\ }\bibfield  {title} {\enquote {\bibinfo {title} {Simplicially driven simple contagion},}\ }\href@noop {} {\bibfield  {journal} {\bibinfo  {journal} {Physical Review Research}\ }\textbf {\bibinfo {volume} {5}},\ \bibinfo {pages} {013201} (\bibinfo {year} {2023})}\BibitemShut {NoStop}%
\bibitem [{\citenamefont {Li}\ \emph {et~al.}(2022)\citenamefont {Li}, \citenamefont {Xue}, \citenamefont {Pan}, \citenamefont {Lin},\ and\ \citenamefont {Wang}}]{li2022competing}%
  \BibitemOpen
  \bibfield  {author} {\bibinfo {author} {\bibfnamefont {W.}~\bibnamefont {Li}}, \bibinfo {author} {\bibfnamefont {X.}~\bibnamefont {Xue}}, \bibinfo {author} {\bibfnamefont {L.}~\bibnamefont {Pan}}, \bibinfo {author} {\bibfnamefont {T.}~\bibnamefont {Lin}}, \ and\ \bibinfo {author} {\bibfnamefont {W.}~\bibnamefont {Wang}},\ }\bibfield  {title} {\enquote {\bibinfo {title} {Competing spreading dynamics in simplicial complex},}\ }\href@noop {} {\bibfield  {journal} {\bibinfo  {journal} {Applied Mathematics and Computation}\ }\textbf {\bibinfo {volume} {412}},\ \bibinfo {pages} {126595} (\bibinfo {year} {2022})}\BibitemShut {NoStop}%
\bibitem [{\citenamefont {Zhao}\ \emph {et~al.}(2024)\citenamefont {Zhao}, \citenamefont {Wang}, \citenamefont {Yang}, \citenamefont {Gu},\ and\ \citenamefont {Moore}}]{zhao2024susceptible}%
  \BibitemOpen
  \bibfield  {author} {\bibinfo {author} {\bibfnamefont {L.}~\bibnamefont {Zhao}}, \bibinfo {author} {\bibfnamefont {H.}~\bibnamefont {Wang}}, \bibinfo {author} {\bibfnamefont {H.}~\bibnamefont {Yang}}, \bibinfo {author} {\bibfnamefont {C.}~\bibnamefont {Gu}}, \ and\ \bibinfo {author} {\bibfnamefont {J.~M.}\ \bibnamefont {Moore}},\ }\bibfield  {title} {\enquote {\bibinfo {title} {Susceptible-infected-recovered-susceptible processes competing on simplicial complexes},}\ }\href@noop {} {\bibfield  {journal} {\bibinfo  {journal} {Physical Review E}\ }\textbf {\bibinfo {volume} {110}},\ \bibinfo {pages} {064311} (\bibinfo {year} {2024})}\BibitemShut {NoStop}%
\bibitem [{\citenamefont {Perc}\ \emph {et~al.}(2013)\citenamefont {Perc}, \citenamefont {G{\'o}mez-Gardenes}, \citenamefont {Szolnoki}, \citenamefont {Flor{\'\i}a},\ and\ \citenamefont {Moreno}}]{perc2013evolutionary}%
  \BibitemOpen
  \bibfield  {author} {\bibinfo {author} {\bibfnamefont {M.}~\bibnamefont {Perc}}, \bibinfo {author} {\bibfnamefont {J.}~\bibnamefont {G{\'o}mez-Gardenes}}, \bibinfo {author} {\bibfnamefont {A.}~\bibnamefont {Szolnoki}}, \bibinfo {author} {\bibfnamefont {L.~M.}\ \bibnamefont {Flor{\'\i}a}}, \ and\ \bibinfo {author} {\bibfnamefont {Y.}~\bibnamefont {Moreno}},\ }\bibfield  {title} {\enquote {\bibinfo {title} {Evolutionary dynamics of group interactions on structured populations: a review},}\ }\href@noop {} {\bibfield  {journal} {\bibinfo  {journal} {Journal of the royal society interface}\ }\textbf {\bibinfo {volume} {10}},\ \bibinfo {pages} {20120997} (\bibinfo {year} {2013})}\BibitemShut {NoStop}%
\bibitem [{\citenamefont {Vendeville}\ and\ \citenamefont {Diaz-Diaz}(2025)}]{vendeville2025modeling}%
  \BibitemOpen
  \bibfield  {author} {\bibinfo {author} {\bibfnamefont {A.}~\bibnamefont {Vendeville}}\ and\ \bibinfo {author} {\bibfnamefont {F.}~\bibnamefont {Diaz-Diaz}},\ }\bibfield  {title} {\enquote {\bibinfo {title} {Modeling echo chamber effects in signed networks},}\ }\href@noop {} {\bibfield  {journal} {\bibinfo  {journal} {Physical Review E}\ }\textbf {\bibinfo {volume} {111}},\ \bibinfo {pages} {024302} (\bibinfo {year} {2025})}\BibitemShut {NoStop}%
\bibitem [{\citenamefont {De~Kemmeter}\ \emph {et~al.}(2024)\citenamefont {De~Kemmeter}, \citenamefont {Gallo}, \citenamefont {Boncoraglio}, \citenamefont {Latora},\ and\ \citenamefont {Carletti}}]{de2024complex}%
  \BibitemOpen
  \bibfield  {author} {\bibinfo {author} {\bibfnamefont {J.-F.}\ \bibnamefont {De~Kemmeter}}, \bibinfo {author} {\bibfnamefont {L.}~\bibnamefont {Gallo}}, \bibinfo {author} {\bibfnamefont {F.}~\bibnamefont {Boncoraglio}}, \bibinfo {author} {\bibfnamefont {V.}~\bibnamefont {Latora}}, \ and\ \bibinfo {author} {\bibfnamefont {T.}~\bibnamefont {Carletti}},\ }\bibfield  {title} {\enquote {\bibinfo {title} {Complex contagion in social systems with distrust},}\ }\href@noop {} {\bibfield  {journal} {\bibinfo  {journal} {Advances in Complex Systems}\ }\textbf {\bibinfo {volume} {2440001}},\ \bibinfo {pages} {2440001} (\bibinfo {year} {2024})}\BibitemShut {NoStop}%
\bibitem [{\citenamefont {Lamata-Ot{\'\i}n}\ \emph {et~al.}(2025)\citenamefont {Lamata-Ot{\'\i}n}, \citenamefont {Malizia}, \citenamefont {Latora}, \citenamefont {Frasca},\ and\ \citenamefont {G{\'o}mez-Garde{\~n}es}}]{lamata2025hyperedge}%
  \BibitemOpen
  \bibfield  {author} {\bibinfo {author} {\bibfnamefont {S.}~\bibnamefont {Lamata-Ot{\'\i}n}}, \bibinfo {author} {\bibfnamefont {F.}~\bibnamefont {Malizia}}, \bibinfo {author} {\bibfnamefont {V.}~\bibnamefont {Latora}}, \bibinfo {author} {\bibfnamefont {M.}~\bibnamefont {Frasca}}, \ and\ \bibinfo {author} {\bibfnamefont {J.}~\bibnamefont {G{\'o}mez-Garde{\~n}es}},\ }\bibfield  {title} {\enquote {\bibinfo {title} {Hyperedge overlap drives synchronizability of systems with higher-order interactions},}\ }\href@noop {} {\bibfield  {journal} {\bibinfo  {journal} {Physical Review E}\ }\textbf {\bibinfo {volume} {111}},\ \bibinfo {pages} {034302} (\bibinfo {year} {2025})}\BibitemShut {NoStop}%
\bibitem [{\citenamefont {Iacopini}, \citenamefont {Karsai},\ and\ \citenamefont {Barrat}(2024)}]{iacopini2024temporal}%
  \BibitemOpen
  \bibfield  {author} {\bibinfo {author} {\bibfnamefont {I.}~\bibnamefont {Iacopini}}, \bibinfo {author} {\bibfnamefont {M.}~\bibnamefont {Karsai}}, \ and\ \bibinfo {author} {\bibfnamefont {A.}~\bibnamefont {Barrat}},\ }\bibfield  {title} {\enquote {\bibinfo {title} {The temporal dynamics of group interactions in higher-order social networks},}\ }\href@noop {} {\bibfield  {journal} {\bibinfo  {journal} {Nature Communications}\ }\textbf {\bibinfo {volume} {15}},\ \bibinfo {pages} {7391} (\bibinfo {year} {2024})}\BibitemShut {NoStop}%
\bibitem [{\citenamefont {Arregui-Garc{\'\i}a}\ \emph {et~al.}(2024)\citenamefont {Arregui-Garc{\'\i}a}, \citenamefont {Longa}, \citenamefont {Lotito}, \citenamefont {Meloni},\ and\ \citenamefont {Cencetti}}]{arregui2024patterns}%
  \BibitemOpen
  \bibfield  {author} {\bibinfo {author} {\bibfnamefont {B.}~\bibnamefont {Arregui-Garc{\'\i}a}}, \bibinfo {author} {\bibfnamefont {A.}~\bibnamefont {Longa}}, \bibinfo {author} {\bibfnamefont {Q.~F.}\ \bibnamefont {Lotito}}, \bibinfo {author} {\bibfnamefont {S.}~\bibnamefont {Meloni}}, \ and\ \bibinfo {author} {\bibfnamefont {G.}~\bibnamefont {Cencetti}},\ }\bibfield  {title} {\enquote {\bibinfo {title} {Patterns in temporal networks with higher-order egocentric structures},}\ }\href@noop {} {\bibfield  {journal} {\bibinfo  {journal} {Entropy}\ }\textbf {\bibinfo {volume} {26}},\ \bibinfo {pages} {256} (\bibinfo {year} {2024})}\BibitemShut {NoStop}%
\bibitem [{\citenamefont {Strogatz}(2018)}]{strogatz2018nonlinear}%
  \BibitemOpen
  \bibfield  {author} {\bibinfo {author} {\bibfnamefont {S.~H.}\ \bibnamefont {Strogatz}},\ }\href@noop {} {\emph {\bibinfo {title} {Nonlinear dynamics and chaos: with applications to physics, biology, chemistry, and engineering}}}\ (\bibinfo  {publisher} {CRC press},\ \bibinfo {year} {2018})\BibitemShut {NoStop}%
\end{thebibliography}%

\renewcommand{\figurename}{Supplementary Fig.}
\renewcommand{\tablename}{Supplementary Table}
\renewcommand{\theequation}{S.\arabic{equation}}

\setcounter{equation}{0}
\setcounter{figure}{0}
\setcounter{section}{0}

\onecolumngrid
\newpage
\section*{Supplementary Information of {\em Genotype networks drive oscillating endemicity and epidemic trajectories in viral evolution}}
\medskip



\section{Validation of the theoretical framework through stochastic simulations}

In Supplementary Fig.~\ref{fig:SM_Fig1}, we compare the stationary dynamics obtained from iterating Eqs.~(1)--(5) with the results of discrete-time stochastic simulations following the rules in Fig.~1. We perform this comparison for each of the cuts shown in Fig.~2, representing the average over 100 independent simulations. In all cases, we observe qualitative agreement between both approaches, which validates the robustness of our theoretical framework. Note that, for the case $\tilde\Delta^{(2)}=0$ (Supplementary Fig.~\ref{fig:SM_Fig1}.a), the discrepancy in the onset of the curves arises due to stochastic fade-out. Moreover, for $\tilde\Delta^{(2)}=0.1$ and $\tilde\Delta^{(2)}=0.15$ (Supplementary Fig.~\ref{fig:SM_Fig1}.b-d), all simulations converge either to the corrupt or the honest stationary state. However, due to the stochastic nature of the simulations, for $\tilde\beta^{(1)}$ values near the thresholds, both outcomes are possible for one set of initial conditions. The criterion we have followed to define the transition in Supplementary Fig.~1 is to consider it occurring when only 10\% of the simulations end up in the minority outcome, assuming these instances arise purely from stochastic effects.

\begin{figure*}[h]
\centering\includegraphics[width=0.86\linewidth]{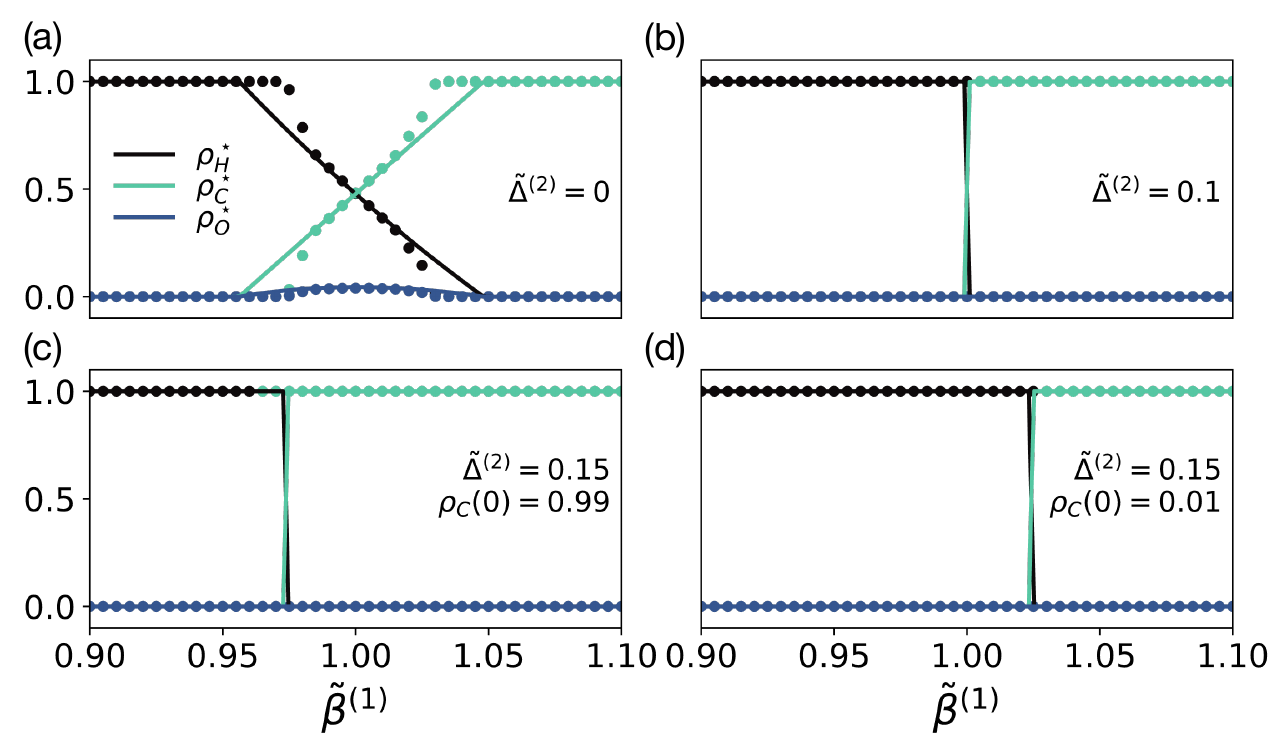}
\caption{\justifying\textbf{Agreement between deterministic dynamics and stochastic simulation}. The three cuts of the diagram in Fig.2 (lines) are compared with the outcome of stochastic simulations (dots). In all panels $k^{(1)}=k^{(2)}=10$, $\mu^{(1)}=0.01$, $\beta^{(2)}=\mu^{(2)}=\Delta/2$ and $r=0.5$. For the stochastic simulations we represent the average over 100 stochastic simulations in a network of $N=2000$ nodes. Note that $\tilde\beta^{(1)}=\beta^{(1)}/\mu^{(1)}$ and $\tilde\Delta^{(2)}=\Delta^{(2)}/\mu^{(1)}$.}
\label{fig:SM_Fig1}
\end{figure*}

\newpage

\section{HO-HCO stability analysis}
\label{sec:apendix1}

In this section, we perform the calculations to derive the stability conditions in Eqs. (6) and (7).

\subsection{Continuous time equations}

We first turn the discrete-time Eqs. (1)-(5) into continuous-time ones. In order to do so, we recall that the probabilities $\beta^{(m)}$ and $\mu^{(m)}$ represent the likelihood of interaction-induced transitions within a given time step. In a continuous-time framework, these are better interpreted as transition rates, leading to exponentially distributed waiting times. Specifically, the discrete-time transition probabilities can be approximated as $\beta^{(m)} \approx 1 - e^{-\hat\beta^{(m)} \, \Delta t} \approx \hat\beta^{(m)} \, \Delta t$, and $\mu^{(m)} \approx 1 - e^{-\hat\mu^{(m)} \, \Delta t} \approx \hat\mu^{(m)} \, \Delta t$,in the limit $\Delta t \to 0$, where $\hat\beta^{(m)}$ and $\hat\mu^{(m)}$ are the continuous-time rates. 

Using this approximation and expanding Eqs.~(1)--(2) to first order in $\Delta t$, we obtain the following differential equations:
\begin{eqnarray}
\dot{\rho}_C(t) &=& -\Pi^{C \to O}(t) \, \rho_C(t) + \Pi^{H \to C}(t) \, \rho_H(t),\label{eq:SM1}\\  
\dot{\rho}_H(t) &=& -\Pi^{H \to C}(t) \, \rho_H(t) + r \, \rho_O(t),\\
\rho_O(t) &=& 1 - \rho_C(t) - \rho_H(t).\label{eq:SM3}
\end{eqnarray}
Now, $\Pi^{H \to C}(t)$ and $\Pi^{C \to O}(t)$ take the form:
\begin{eqnarray}
\Pi^{H \to C}(t) &=& 1 - \prod_{m=1}^M \left( 1 - \hat\beta^{(m)} \, \rho_C^{(m)}(t) \right)^{k^{(m)}},\\
\Pi^{C \to O}(t) &=& 1 - \prod_{m=1}^M \left( 1 - \hat\mu^{(m)} \, \rho_H^{(m)}(t) \right)^{k^{(m)}},
\end{eqnarray}
with the understanding that the rates $\hat\beta^{(m)}$ and $\hat\mu^{(m)}$ capture the instantaneous likelihood per unit time of a successful contagion or delation event, respectively.

\subsection{Derivation of the stability condition}

To determine the stability conditions of the full-honesty and full-corrupt states, here we construct the Jacobian matrix of the set of Eqs. (\ref{eq:SM1})-(\ref{eq:SM3}). The Jacobian matrix reads:
\renewcommand{\arraystretch}{1.5} 
\begin{eqnarray}
J = \begin{bmatrix}
\frac{\partial \dot\rho_H}{\partial \rho_H} & \frac{\partial \dot\rho_H}{\partial \rho_C} \\
\frac{\partial \dot\rho_C}{\partial \rho_H} & \frac{\partial \dot\rho_C}{\partial \rho_C} \\
\end{bmatrix},
\end{eqnarray}
where the partial derivatives take the following expression 
\begin{eqnarray}
    \frac{\partial \dot\rho_H}{\partial \rho_H}&=&-r-\Pi^{H\rightarrow C},\\
    \frac{\partial \dot\rho_H}{\partial \rho_C}&=&-r-\rho_H\sum_{m=1}^M\left\{mk^{(m)}\beta^{(m)}\rho_C^{m-1}\prod_{n=1}^M\left[\frac{\left[1-\beta^{(n)}\rho_C^n\right]^{k^{(n)}}}{\left[1-\beta^{(m)}\rho_C^m\right]}\right]\right\},\\
    \frac{\partial \dot\rho_C}{\partial \rho_H}&=&\Pi^{H\rightarrow C}-\rho_C\sum_{m=1}^M\left\{mk^{(m)}\mu^{(m)}\rho_H^{m-1}\prod_{n=1}^M\left[\frac{\left[1-\mu^{(n)}\rho_H^n\right]^{k^{(n)}}}{\left[1-\mu^{(m)}\rho_H^m\right]}\right]\right\},\\
    \frac{\partial \dot\rho_C}{\partial \rho_C}&=&\rho_H\sum_{m=1}^M\left\{mk^{(m)}\beta^{(m)}\rho_C^{m-1}\prod_{n=1}^M\left[\frac{\left[1-\beta^{(n)}\rho_C^n\right]^{k^{(n)}}}{\left[1-\beta^{(m)}\rho_C^m\right]}\right]\right\}-\Pi^{C\rightarrow O}.
\end{eqnarray}
Note that, for simplicity, we refer as $\beta^{(m)}$ and $\mu^{(m)}$ to the instantaneous likelihoods per unit time of a successful event. When evaluated in the full-honesty state ($\rho_H=1,\rho_C=0$), the Jacobian matrix becomes
\begin{eqnarray}
J_H = \begin{bmatrix}
-r & -r-k^{(1)}\beta^{(1)} \\
0 & k^{(1)}\beta^{(1)}-1+\prod_{m=1}^M\left[1-\mu^{(m)}\right]^{k^{(m)}} \\
\end{bmatrix},
\end{eqnarray}
and in the full-corrupt ($\rho_H=0,\rho_C=1$), the Jacobian matrix reads
\begin{eqnarray}
J_C = \begin{bmatrix}
-r -1 +\prod_{m=1}^M\left[1-\beta^{(m)}\right]^{k^{(m)}}  & -r \\
-k^{(1)}\mu^{(1)}+1-\prod_{m=1}^M\left[1-\beta^{(m)}\right]^{k^{(m)}}  & 0 \\
\end{bmatrix}.
\end{eqnarray}
The eigenvalues of the jacobians are the roots of the characteristic polynomial $\lambda^2-\lambda T+D$, where $T=Tr(J)$ and $D=Det(J)$ are respectively the trace and the determinant of the Jacobian matrices. Considering that in both cases $T<0$, the stability condition requires that $D>0$ \cite{strogatz2018nonlinear}. Explicitly, the stability condition of the full-honesty state reads
\begin{eqnarray}
    k^{(1)}\beta^{(1)}-1+\prod_{m=1}^M\left[1-\mu^{(m)}\right]^{k^{(m)}}<0,
\end{eqnarray}
while the condition for the full-corrupt state is
\begin{eqnarray}
    k^{(1)}\mu^{(1)}-1+\prod_{m=1}^M\left[1-\beta^{(m)}\right]^{k^{(m)}}<0.
\end{eqnarray}

\section{Critical points derivation}
\label{sec:apendix2}

To obtain Eq. (19) we have imposed that $\rho^{\star}=0$ in Eq. (18). Moreover, to obtain Eq. (20) we proceed as follows. Eq. (17) can be simplified by introduced $\lambda^{(m)}$ and $\gamma^{(m)}$ and rescaling the time by $\mu^{(0)}$. Moreover, imposing stationary conditions ($\dot\rho=0$), the expression reads:
\begin{eqnarray}
    0&=&\lambda^{(1)}\rho^{\star}(1-\rho^{\star})+\lambda^{(2)}\rho^{\star\,2}(1-\rho^{\star})\nonumber \\
    &&-\gamma^{(1)}(1-\rho^{\star})\rho^{\star}-\gamma^{(2)}(1-\rho^{\star})^2\rho^{\star}-\rho^{\star}.
\end{eqnarray}
Discarding the absorbent solution $\rho^{\star}=0$ and rearranging terms we reach
\begin{eqnarray}
    0&=&\rho^{\star\,2}\left(\lambda^{(2)}+\gamma^{(2)}\right)\nonumber\\
    &&+\rho\left(\lambda^{(1)}-\gamma^{(1)}-\lambda^{(2)}-2\gamma^{(2)}\right)\nonumber\\
    &&+1+\gamma^{(1)}+\gamma^{(2)}-\lambda^{(1)}.
    \label{eq:B2}
\end{eqnarray}
From this latter equation, Eq. (18) can be easily derived. However, to obtain the critical value of $\lambda^{(1)}$ determining the onset of the transition, we need to get the values ($\lambda,\rho^{\star}$) such that $\rho^{\star}=\rho^{\star}_{2\,+}=\rho^{\star}_{2\,-}$. This occurs at the saddle point ($\lambda_{c\,2},\rho^{\star}_{c}$) depicted in Fig. 4.a-b, that fulfills that $\left(\frac{\partial \lambda^{(1)}}{\partial \rho^{\star}}\right)_{\lambda_{c\,2},\rho^{\star}_{c}}=0$. From Eq. (\ref{eq:B2}) we clear that
\begin{eqnarray}
    \lambda^{(1)}&=&\frac{\rho^{\star\,2}\left(\lambda^{(2)}+\gamma^{(2)}\right)-\rho^{\star}\left(\gamma^{(1)}+\lambda^{(2)}+2\gamma^{(2)}\right)+1+\gamma^{(1)}+\gamma^{(2)}}{1-\rho^{\star}},
\end{eqnarray}
and by imposing that $\left(\frac{\partial \lambda^{(1)}}{\partial \rho^{\star}}\right)_{\lambda_{c,2}\,\rho^{\star}_{c}}=0$ we obtain after some algebra:
\begin{eqnarray}
    \rho^{\star}_c&=&\frac{\left(\lambda^{(2)}+\gamma^{(2)}\right)\pm\sqrt{\lambda^{(2)}+\gamma^{(2)}}}{\left(\lambda^{(2)}+\gamma^{(2)}\right)},\\
    \lambda^{(1)}_{c,\,2}&=&2\sqrt{\lambda^{(2)}+\gamma^{(2)}}+\gamma^{(1)}-\lambda^{(2)}.
\end{eqnarray}
Furthermore, we can derive the onset of explosivity by setting $\rho^{\star}_c=0$, which leads to the condition in Eq. (21).

\newpage

\section{Bifurcation diagram of behavior adoption dynamics}

In Supplementary Fig. \ref{fig:SM_Fig2} we separate Fig. 5 in two different panels: Supplementary Fig. \ref{fig:SM_Fig2}.a showing the dependency of $\rho_c^{\star}$, and \ref{fig:SM_Fig2}.b showing the dependency of $\lambda_{c,\,2}^{(1)}$.

\begin{figure}[h]
\centering\includegraphics[width=1\linewidth]{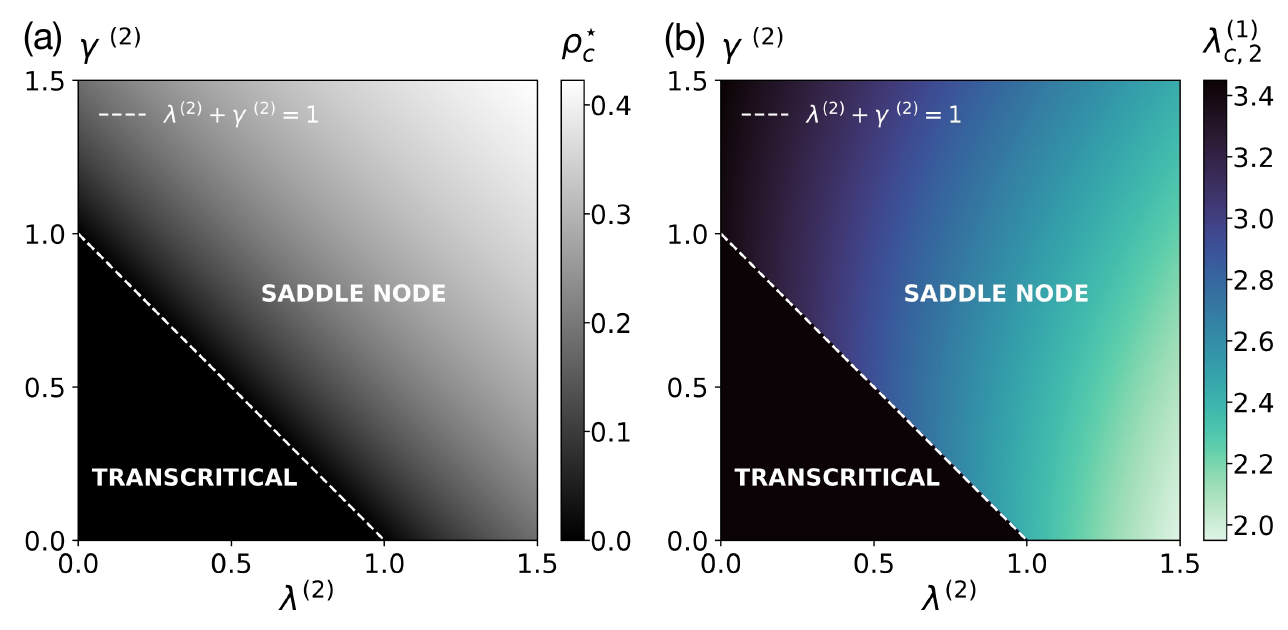}
\caption{\justifying\textbf{Bifurcation diagram of behavior adoption dynamics}. Color code corresponds to the critical point ($\rho_{2\,+}^{\star}$ in panel (a) and $\lambda^{(1)}_{c,\,2}$ in panel (b)) in terms of the higher-order adoption parameters $\lambda^{(2)}$ and $\lambda^{(2)}$. As the relevancy of higher-order interaction increases, the local bifurcation shifts from transcritical to saddle-node. The dashed line indicates the boundary. Note that we have set $\gamma^{(1)}=1$.}
\label{fig:SM_Fig2}
\end{figure}

\newpage

\end{document}